%% file: Brick_filaments.tex
\documentclass[12pt,preprint]{aastex}

\usepackage{rotating}

\newcommand{\cmq}{cm{$^{-3}$}}

\newcommand{\msol}{M{$_{\odot}$}}

\newcommand{\kms}{km~s{$^{-1}$}}

\newcommand{\Vlsr}{V{$_{LSR}$}}

\newcommand{\HCOP}{{HCO$^+$}}
\newcommand{\hcop}{{H$^{13}$CO$^+$}}
\newcommand{\hcn}{{H$^{13}$CN}}

\slugcomment{DRAFT: \today}
\shorttitle{Brick Filaments}
\shortauthors{Bally et al.}

\begin{document}

\title{Absorption Filaments Towards the Massive Clump G0.253+0.016}

\author{
               John Bally \altaffilmark{1},
               J. M. Rathborne \altaffilmark{2},
               S, N. Longmore \altaffilmark{3}, 
               J. M. Jackson \altaffilmark{4},
               J. F. Alves \altaffilmark{5}, 
                E.  Bressert \altaffilmark{6}, 
               Y.  Contreras  \altaffilmark{7},
                J. B. Foster \altaffilmark{8}, 
               G. Garay \altaffilmark{9},
               A. Ginsburg \altaffilmark{10},
               K. G. Johnston \altaffilmark{11},
               J. M. D. Kruijssen \altaffilmark{12},
                L. Testi \altaffilmark{13},
                    and
               A. J. Walsh \altaffilmark{14}
               	    }
	    
\affil{{$^1$}{\it{
 Department of Astrophysical and Planetary Sciences,\\
 University of Colorado, UCB 389, 
 Boulder, CO 80309}      
 \email{John.Bally@colorado.edu}}  }
          
 \affil{{$^2$}{\it{
 CSIRO Astronomy and Space Science, 
 P.O. Box 76, Epping NSW, 1710, Australia }
 \email{Jill.Rathborne@csiro.au}}   }
 
  \affil{{$^{3}$}{\it{
 Liverpool Johns Moore University,  
 Liverpool, England }
 \email{S.N.Longmore@ljmu.ac.uk}}    }
 
 \affil{{$^{4}$}{\it{
 Institute for Astrophysical Research,
 Boston University, Boston, MA 02215, USA} 
 \email{jackson@bu.edu}}   }

  \affil{{$^5$}{\it{
  University of Vienna, 
  Turkenschanzstrasse 17, 1180 Vienna, Austria  } 
 \email{joao.alves@univie.ac.at}}  }

 \affil{{$^6$}{\it{
 CSIRO Astronomy and Space Science, 
 P.O. Box 76, Epping NSW, 1710, Austraila }
 \email{ebressert@gmail.com}}   }    
 
  \affil{{$^7$}{\it{
 CSIRO Astronomy and Space Science, 
 P.O. Box 76, Epping NSW, 1710, Austraila }
 \email{Yanett.Contreras@csiro.au }}   }    
 
  \affil{{$^8$}{\it{
  Yale Center for Astronomy and Astrophysics, Yale University, 
  New Haven, CT 06520, USA  } 
  \email{jonathan.b.foster@yale.edu}} }

 \affil{{$^{9}$}{\it{
  Universidad de Chile, 
  Camino El Observatorio 515, Las Condes, Santiago, Chile } 
  \email{guido@das.uchile.cl}} } 

 \affil{{$^{10}$}{\it{
 European Southern Observatory, 
 Karl-Schwarzschild-Str. 2, 85748 Garching bei Munchen, Germany }
 \email{adam.g.ginsburg@gmail.com}} }
 
  \affil{{$^{11}$}{\it{Max Planck Institute for Astronomy,
 Heidelberg, Germany,  \\
 School of Physics and Astronomy, University of Leeds, Leeds LS2 9JT, UK
 }
 \email{johnston@mpia-hd.mpg.de}} }
 
  \affil{{$^{12}$}{\it{
 Max Planck Institut fur Astrophysik, 
 Karl-Schwarzschild-Str. 2,  Garching bei Munchen, Germany }
 \email{kruijssen@mpa-garching.mpg.de}}  }
 
  \affil{{$^{13}$}{\it{
 European Southern Observatory, 
 Karl-Schwarzschild-Str. 2, 85748 Garching bei Munchen, Germany }
 \email{ltesti@eso.org}} }

 \affil{{$^{14}$}{\it{
 International Centre for Radio Astronomy Research, 
 Curtin University, GPO Box U1987, Perth, Australia }
 \email{andrew.walsh@curtin.edu.au}} }
   
\begin{abstract}
ALMA \HCOP\ observations of the infrared dark cloud
G0.253+0.016  located in the Central Molecular Zone 
of the Galaxy are presented.  The 89~GHz emission is 
area-filling, optically thick, and  sub-thermally excited.
Two types of filaments are seen in absorption against the 
\HCOP\ emission.
{\it Broad-line absorption filaments} (BLAs) have widths of 
less than a few arcseconds (0.07 $-$ 0.14 pc), lengths of 
30 to 50 arcseconds  (1.2 $-$1.8 pc), and absorption profiles 
extending  over a velocity range larger than 20 \kms .  The 
BLAs are nearly parallel to the nearby G0.18 non-thermal filaments  
and may trace \HCOP\  molecules gyrating  about highly ordered 
magnetic fields located in front of G0.253+0.016 or edge-on sheets  
formed behind supersonic shocks propagating orthogonal 
to our line-of-sight in the foreground. 
{\it Narrow-line absorption filaments} (NLAs) 
have line-widths less  than 20 \kms .
Some NLAs  are also seen in  absorption  in other  species  
with high optical depth such as HCN and occasionally in 
emission where the background is faint.  The NLAs, which also 
trace low-density, sub-thermally excited \HCOP\ molecules,  
are mostly seen on the  blueshifted side of the emission 
from G0.253+0.016.    If associated with the surface of 
G0.253+0.016, the kinematics of the NLAs  indicate  that the 
cloud surface is expanding.   The decompression of entrained, 
milli-Gauss magnetic fields may be responsible for the 
re-expansion of the surface layers of G0.253+0.016  
as it recedes from the Galactic center following a close 
encounter with Sgr A.  
\end{abstract}


\section{Introduction}

The Central Molecular Zone (CMZ) at Galactocentric radii less 
than $\sim$500 pc contains the most massive, dense, and turbulent 
molecular clouds in the Galaxy 
\citep{RodriguezFernandez2008,Ferriere2007, RodriguezFernandez2006,
Pierce-Price2000, MorrisSerabyn1996, BallyGC1987,BallyGC1988}
along with some of the most compact and massive star clusters 
\citep{Figer2002,Habibi2013}.  

\citet{LisCarlstrom1994}
found an exceptionally massive and  dense cloud seen in projection against 
the bright near- and mid-infrared emission from the CMZ, G0.253+0.016.   
Located towards the brightest portion of the Galactic plane at
infrared wavelengths,  this cloud was clearly seen in silhouette 
against the CMZ  in the IRAS mid-infrared images and is the most 
prominent infrared dark cloud (IRDC) in the sky 
\citep{Menten_etal2005,Arendt2008,Ramirez2008}.   

\citet{Lis_etal_1994} and \citet{LisMenten1998} demonstrated that this cloud has an 
unusually high mass ($M > 10^5$ \msol ) and 
density ($N(H_2) > 10^5$ cm$^{-2}$) and is very compact ($r < $ 3 pc).    
\citet{Lis2001} used ISO data and millimeter-wave spectra to show that 
the dust temperature is low but the gas is hot.   In the cloud center,  
15 $<~T_{dust} <$ 22 K  but increases to $\sim$27 K at the cloud edges
\citep{Longmore2012}.   On the other hand, the gas is hot with 
$T_{gas} \sim$ 80 K \citep{Ao2013}.  The  mid-infrared fine-structure cooling
lines measured with ISO indicated that near-UV interstellar 
radiation field surrounding the cloud is about  $10^3$ times stronger than in the
Solar vicinity \citep{Lis2001}, consistent with warmer dust at the cloud periphery. 
\citet{Kauffmann2013} presented interferometric SiO and  N$_2$H$^+$ maps 
of G0.253+0.01 that exhibit some localized narrow spectral lines
with central velocities covering the wide velocity range seen in single-dish
data \citep{Rathborne2014_MALT90}.   
The large line-widths and bright emission in tracers such
as SiO, and the complicated kinematics was interpreted as evidence for
shocks, possibly indicating an early stage of a cloud-cloud collision
\citep{Higuchi2014}.

G0.253+0.016  (also known as the `Lima Bean' or the `Brick') has been 
proposed to be a possible progenitor to a massive star cluster 
\citep{Longmore2012}.
Fitting the dust continuum SED to the Herschel 60, 160, 250, 350, and 
500 $\mu$m data obtained  from the Herschel Space Observatory 
Hi-GAL survey  \citep{Molinari2011, Molinari2010a,Molinari2010b} 
and 1100 $\mu$m from  the Bolocam Galactic Plane Survey (BGPS; Bally et al. 2010) 
reveals a dust  temperature of  $\sim$20 K, H$_2$ column density of  
$\sim 3.3 \times 10^{23}$~cm$^{-2}$, and mass of  $\sim 10^5$ \msol\ 
\citep{Longmore2012,Longmore2013}.   Absorption in the near- and far-infrared 
and application of the virial theorem to spectral line data {\citep{Rathborne2014_MALT90}
give similar results for the mass and column density.   With a mean radius of 
about 3 pc  and a  mean density $n(H_2) \approx 7 \times 10^{4}$ cm$^{-3}$, 
G0.253+0.016 has sufficient mass to potentially 
form a young massive cluster \citep{PortegiesZwart2010}.  
However, previous studies found no evidence for 
internal heating sources  or embedded HII regions \citep{Lis2001}.  
Only one H$_2$O maser has been found  in G0.253+0.016
\citep{Lis_etal_1994,BreenEllingsen2011}.    Thus, G0.253+0.016 may either
be in the very  earliest stages of high-mass star  or cluster formation, or because
of the extreme conditions in the CMZ, may fail to form many stars.
 
Most star formation in the CMZ occurs on a twisted elliptical ring-like
structure with a projected radius of about 100 pc and offset towards positive 
Galactic longitudes with respect to the dynamical center of the Galaxy 
\citep{RodriguezFernandez2006,Molinari2011}.  The ring contains the Sgr C, 
Sgr A (which contains the supermassive black hole marking the 
center of the Galaxy), Sgr B1,  and Sgr B2   star forming complexes.  
G0.253+0.016 is the  most prominent member of a chain of IRDCs  at 
positive Galactic longitudes between Sgr A and Sgr B2 
\citep{Lis2001,Longmore2013}.    \citet{Longmore2012} and \citet{Rathborne2014_MALT90}
 found that G0.253+0.016 exhibits a line-width near zero intensity of 
 about $\sim$50 to 60 \kms ,  and an internal velocity dispersion of  at least 16 \kms\
 typical for CMZ clouds that  implies a crossing time of about  0.17 Myr.  
  
 Models of orbits in barred, tri-axial potentials  indicate that gas in the inner 
 CMZ is likely to be moving on  $x_2$ orbits  elongated along the minor axis of the bar
 \citep{Contopoulos1980,Athanassoula1992b,Athanassoula1992a}.
 The major axis of the bar is oriented between 20 and 45 degrees with
 respect to our line of sight \citep{Binney1991,RodriguezFernandez2008}.   
 G0.253+0.016  is located at a projected distance of
 about 50 pc from Sgr A  and has a radial velocity \Vlsr\  $\approx$  40 \kms. 
 Assuming that it is moving close to the plane of the
 sky with a velocity  of about 150 to 180 \kms\  along an elongated $x_2$ orbit 
 (this is the speed expected for a circular orbit given the enclosed mass), it 
 would have passed near Sgr A about 0.25 to 0.3 Myr ago \citep{Kruijssen2014}.
 Clouds injected onto $x_2$ orbits near apocenter  with velocities lower than
 the circular orbit speed at that location tend to plunge deeper into the 
 potential well.  Following pericenter passage near Sgr A, they climb out of the 
 potential to a second apocenter  located on the opposite side of the nucleus,  and 
 on the opposite side of the Galaxy from the first apocenter.   
 Such plunging, elongated  orbits  precess at rates which depend on the details 
 of the enclosed mass distribution and orbit eccentricity.    
 \citet{Longmore2013} proposed that the encounter of  G0.253+0.016 with
 Sgr A  may have compressed this cloud.  
  
 In this paper, high-angular resolution, high dynamic range, and sensitive
  ALMA observations of  J = 1$-$0 \HCOP\  are presented that
 reveal an extensive network of  filaments seen in absorption toward G0.253+0.016.   
 The most prominent filament (Figure \ref{fig1}) has an absorption line-width of over 
 30 \kms\ and is nearly parallel to the bright, non-thermal radio filaments which cross 
 the Galactic plane at longitude 0.18\arcdeg\ a few arc minutes from G0.253+0.016.  
 Dozens of filaments with less than 20 \kms\  linewidths are predominantly seen on the 
 blueshifted side of the \HCOP\ emission from G0.253+0.016.   In this paper, it is shown
 that these absorption features are produced by  very low-density, sub-thermally 
 excited \HCOP\ absorbing optically thick, but nevertheless subthermaly excited
 background \HCOP\ emission from this cloud.   If  located on the near 
 surface of G0.253+0.016, their blueshifted velocities with respect to the background
 \HCOP\ emission would indicate that the cloud surface layers are expanding,  
 contrary to the naive expectation of global collapse if it were a progenitor 
 to a massive cluster. 

\section{Observations}

Observations presented here were obtained with the Atacama Large Millimeter Array
(ALMA) located on the Chajnantor plateau in the Northern Chilean
Andes.   Full details of the observations and data reduction will be presented in another
paper  \citep{Rathborne2014_ALMA}.    A short summary is presented below.

G0.253+0.016  was imaged in the 3 mm wavelength 
bands (ALMA band 3)  using 25 of ALMA's 12-meter diameter antennas.
The synthesized beam produced by ALMA at this frequency 
has a diameter of 1.7\arcsec\ corresponding
to a linear scale of  0.07 pc at the 8.5 kpc distance of the CMZ.  
Because the primary-beam field-of-view of ALMA is small ($\sim$ 70\arcsec)
and G0.253+0.016 subtends a 1\arcmin\  by 3\arcmin\ region, a mosaic consisting of
13 separate pointings was required.  Baselines ranged from 13 to 455 meters.
G0.253+0.016 was imaged on six separate occasions over the period 29 July to 1 
August 2012.    Each data set was independently calibrated prior to merging.
The ALMA correlator was configured to measure the continuum in a 1.875 GHz 
bandpass, and to observe 10 different molecular emission lines with a resolution of 
0.488 MHz (about 1.5 \kms\ velocity resolution) covering a velocity range from 
about \Vlsr\ = -200 \kms\   to about +240 \kms .   
Single-dish observations of G0.253+0.016 in the same transitions as observed with
ALMA were obtained with the 22 meter Mopra radio telescope in Australia as part of
the MALT90 Survey \citep{Foster2011,Jackson2013,Foster2013}.  
The resulting maps were used to restore 0-spacings in the ALMA data in the
Fourier domain \citep{Rathborne2014_ALMA}.   

Final data cubes were gridded to a pixel scale of 0.35\arcsec\ per pixel.
Because the flux in adjacent channels in the data is
correlated, the data were re-sampled to 3.4 \kms\  per channel.  
The rms noise in each frequency channel is about 5 mJy (0.26 K) in the 
1.7\arcsec\  beam.   At the  rest frequency of the HCO$^+$ J = 1-0 transition
(89.1885 GHz)  a flux of 1 Jy in the  1.7\arcsec\  diameter beam (or
a surface brightness of 1 Jy/beam)  corresponds to a brightness 
temperature of 53.5 K.     All temperatures and fluxes refer to values above
the cosmic microwave background or any other smooth background 
which are resolved out by ALMA and  removed from the single-dish 
observations by beam-switching.

\section{Results}

The blueshifted emission in  \HCOP\  shows an extensive network of 
curved and linear filamentary structures seen in absorption against the warmer 
background  emission  from G0.253+0.016.   Figure \ref{fig1} shows an integrated 
intensity map of G0.253+0.016 in the J = 1-0 \HCOP\  transition.   While the 
ALMA \HCOP\ integrated intensity map is dominated by area-filling emission,
the optically thinner  \hcop\ cubes show bright arcs resembling bows moving towards 
increasing R.A. (left in the figures).  This overall morphology is also seen in other 
optically thin (or at least thinner than \HCOP) tracers such as HNCO, H$_2$CS, 
SiO, and SO  \citep{Rathborne2014_ALMA}.    

Figures 2 through 9 show  individual 3.4 \kms\ wide channel maps in \HCOP . 
The yellow arrows mark the location of a pair of  filaments having 
a length of over $L$ $>$ 60\arcsec\  and a radial velocity dispersion greater than
20 \kms .    Dozens of filaments with line-widths less than 10 to 20 \kms\  are located
on the blueshifted (low velocity) side of the \HCOP\  emission.     These 
filaments also have widths of only a few arcseconds,  but their lengths 
tend to be shorter  than the broad-line filaments.   Some filaments form clusters of
nearly parallel strands  which at  extreme blueshifted radial velocities  
(\Vlsr\ = 5 to 30 \kms )   blend into  a complex network of absorption covering most 
of the spatial extent of G0.253+0.016.   The morphology of the absorption filaments 
do not show any correlation with the  underlying  filamentary structure of 
G0.253+0.016 seen in optically thin emission lines.
The absorption filaments can be subdivided into two categories based 
on their observed properties:

{\it Broad-line absorption (BLA)} filaments have widths of less than a few 
arcseconds  ($<$ 0.04 to 0.08 pc),  lengths of 30 to 50 arcseconds 
($\sim$ 1 to 2 pc),   and absorption profiles  extending over a velocity  
range larger than 20 \kms\  around  \Vlsr\  = 10 to 40 \kms  .   The two most 
evident BLAs (marked by in Figures 1  through 9)    are nearly 
parallel to the nearby  G0.18 non-thermal filaments seen in 20 and 6 cm 
wavelength  radio continuum images of  \citet{Yusef-Zadeh2004}.   These  BLAs 
have position angles (measured from North to East) of PA $\sim$ 125\arcdeg\ to
140\arcdeg , nearly perpendicular to the Galactic plane.  The nearby non-thermal 
filaments have PA $\sim$ 125\arcdeg . 

Analysis of spatial-velocity cuts and spectra orthogonal to the most prominent  
BLA filament in Figure 1 shows that it extends from below 17 \kms\ to about 
45 \kms\   (Figure \ref{fig10}).    
The intensity in the faintest part  of the filament ranges from about 0.17 to 
greater than 0.8   times the \HCOP\ intensity in the surrounding
region which has a brightness temperature of 3 to 4 K (0.05 to 0.08 Jy/beam).   
Thus corresponding brightness temperatures at the most opaque portions
of the BLA filaments range from $\sim$ 0.5 K (0.01 Jy/beam) at the darkest part 
of the main BLA to  about  2 K (0.04 Jy/beam) at more typical locations. 
The lowest-opacity occurs near the ends of the major BLA filaments.  
The deepest absorption occurs at the redshifted end of the feature near
\Vlsr\ $\approx$ 37 \kms\ at J2000 $\approx$ 17:46:09.79, $-$28:42:03.5.  
The mean flux  along cross-cuts is about 0.5 times the surrounding cloud flux.   
Thus, the optical depth of the absorbing gas ranges from $\sim$ 0.2 to about 
2 with a mean value of  0.7. The  \HCOP\ BLA filaments have no 
counterparts in any of the other spectral lines.   Specifically, they are not 
seen  in other optically thick species  such as HCN.

{\it Narrow-line absorption (NLA)} \HCOP\  filaments, some of which are also 
seen in absorption in other  species with high optical depth such as HCN,   
have line widths of less than 20 \kms .    Some tend to cluster into filament 
bundles which taken together have velocity extents up to 20 \kms .    These 
filaments  are mostly located at   at \Vlsr\ $\sim$ 10 to 40 \kms,  blueshifted 
relative to the median radial velocity of  G0.253+0.016.

Table~1 lists several examples of NLA filaments.   Near the Northwest portion 
of the cloud, several arcs form nearly  concentric loops 
with radii of about 6\arcsec\  and 13\arcsec\  between \Vlsr\ = 11 and 21 
\kms\  (Figures \ref{fig2}  and \ref{fig3}).   NLA1 is  a narrow line-width
absorption feature about 20\arcsec\ south of the two BLAs.  
The NLA2  consists of several parallel east-west strands near the middle of 
the cloud around $\delta$ = -28:42:50 
(Figures \ref{fig5}, \ref{fig6}, and  \ref{fig7}).   
NLA 3  is a cluster of bent filaments  near the redshifted
southern part of G0.253+0.016 around $\delta$ = -28:43:00 to - 28:44:00
(Figures \ref{fig4}, \ref{fig5}, and  \ref{fig6}).     Many  NLA filaments
blend into broad regions of absorption on the blueshifted, low-radial
velocity side of the \HCOP\ data cube.   
Figure \ref{fig11} shows spectra of an  NLA filament near the northern
side of the NLA3 cluster.
Figure \ref{fig12} shows the \HCOP\ integrated intensity map of  G0.253+0.016
with the locations of  the BLA and NLA filaments with the locations of
the features listed in Table~1 marked.

At the extreme blueshifted end of the data cube, the NLA filament network
blends into an area-filling absorption.   
The channel maps between \Vlsr\ = 14 and 21 \kms\ indicate that most of
the surface of G0.253+0.016 is covered by NLA filaments.    The fractional 
area covered diminishes toward higher radial velocities.
Most NLAs are shorter than the BLAs and many
are curved.  A few, such as those centered near 
J2000 = 17:46:10.6, -28:41:55   trace nearly complete circles.  
Others are partial arcs.   The distinction between  BLA  and NLA filaments 
is not sharp;  they form a continuum with decreasing 
line-widths and lengths from BLA1 and 2 (Table~1)  to the narrower and shorter
NLA filaments.

Comparison with the HCN  and HNCO data (not shown) 
indicate that some  \HCOP\ NLAs are also seen as absorption 
features in these tracers (Figure \ref{fig13}).    
A few \HCOP\ NLA segments go into emission where their 
surroundings are dim in  HCN or HNCO  such as near the projected edges of 
the cloud.   These filaments are  confined to within a few arc-seconds of the 
cloud's edge,  indicating that they may be close to  the
surface of G0.253+0.016.   
 
The distribution of \HCOP\ absorption features  have 
no correlation with  the underlying structure of G0.253+0.016 seen in 
other tracers.    G0.253+0.016 is filamentary in optically thin emission 
lines such as  \hcn\  where  G0.253+0.016 resembles a bow shock propagating 
towards the east.   Figure \ref{fig14} shows the optically thin \hcop\ in three
broad velocity ranges 
V$_{LSR}$ = 10.8 to 27.8 \kms\ (blue),  31.2 to 41.4 \kms\ (green), and
44.8 to 61.8 \kms\ (red).   
The east-facing bow shapes are not visible in the optically thick \HCOP\  
(or  HCN) data presumably because this emission is area-filling and these 
tracers only  probe cloud structure to the $\tau \approx$ 1 surface. 
The  emission-line structure of G0.253+0.016  is discussed
in \citet{Rathborne2014_ALMA}.  For reference, Figure \ref{fig15} shows the 
3 mm continuum emission from \citet{Rathborne2014_ALMA}.    It is unclear what 
fraction of this emission is produced by dust and what fraction is produced
by free-free emission.

\subsection{Opacity and Excitation Conditions for  \HCOP }

The typical  \HCOP\ brightness temperature in both the ALMA  and 
single dish  observations is only  about 2 to 4 K 
\citep{Rathborne2014_ALMA, Rathborne2014_MALT90},  
considerably  lower than the gas temperatures of 60 to 80 K inferred from 
NH$_3$ or H$_2$CO  observations \citep{Lis2001,MillsMorris2013,Ao2013} 
and even lower  than the dust temperature, which  has 
been estimated to be around  19 K in the  cloud center and about 27 K 
at the  cloud surface  \citep{Lis2001,Pierce-Price2000,Longmore2012}. 
The ALMA data shows that the \HCOP\ emission is area filling at arcsecond
scales.    Thus, the low brightness temperature must be due to low excitation
temperature at the $\tau_{12}$ = 1 surface, implying densities much lower 
than the critical density for the J=1$-$0  89 GHz transition of \HCOP\ 
(for which the critical density is  $\sim 2 \times 10^5$ \cmq ).

Comparison with the \hcop\  ALMA images  demonstrates that the \HCOP\ 
emission is optically  thick at most locations.
The area-integrated intensity ratio of \HCOP\ divided by the area-integrated  
ratio of \hcop\   is $R_{12/13}$ =  I(\HCOP )/I(\hcop ) = $10\pm 1$ for the 
northern part of G0.253+0.016 and $R_{12/13} = 11\pm 1$ for the southern part
for integration areas ranging from 0.3 to 2.1 square arc-minutes.    
The maximum \HCOP\ brightness (0.23 Jy/beam or 12.3 Kelvin) occurs  
in a small  3\arcsec\ by 8\arcsec\ elongated knot  near the southern part of 
G0.253+0.016 at  J2000 = 17:46:08.5, $-$28:43:33  at \Vlsr\ = 55 \kms .   
At this location,  the  ratio $R_{12/13} \approx 12$.    Thus, the observed
brightness ratio,   $R_{12/13}$ is lower than the abundance ratio,
$X_{12/13}$ =  $[^{12}C] / [^{13}C]$.

\citet{LangerPenzias1990} measured the abundance ratio $X_{12/13}$    
as a function of  Galactocentric radius  finding that in the CMZ  
$X_{12/13} \approx $ 24.
In a more recent study, \citet{ Riquelme2010} found a similar ratio in most
parts of the CMZ.  However, near cloud edges, some  ratios as  high as 
40 were found. \citet{Milam2005}  found a lower value in the CMZ of 
$X_{12/13} \approx $ 17 $\pm$ 7.    Thus, there may a some intrinsic
variation possibly do to local enrichments or depletion such as might
be caused by local stellar processing and injection into the ISM 
or isotope-selective photo-dissociation
\citep{RolligOssenkopf2013,LyonsYoung2005, BallyLanger1982}.  

Assuming that the $X_{12/13}$  is 24  in G0.253+0.016, 
that  the excitation temperatures of \hcop\ and \HCOP\ are
the same, and that for \HCOP\ the observed brightness is equal to 
the excitation temperature, as expected for an optically thick emission line,
the  $R_{12/13}$ ratio  of  $\sim$10 indicates that the mean emission  line
optical depth of \HCOP\  is
$$
\tau _{12} \approx   X_{12/13} ln [R_{12/13} / (R_{12/13} - 1)] \sim 2.5
$$   
For $X_{12/13}$ ranging from 17 to 40, $\tau _{12}$, $R_{12/13}$ = 10 
implies that $\tau_{12}$ ranges from  1.5 to 2.5.  

In the \HCOP\  photosphere (where $\tau_{12} \sim$ 1), the density must be 
much lower than the critical density, implying sub-thermal excitation, and 
densities much lower than the mean density of G0.253+0.016.   The brightness
temperature of  \HCOP\  implies that $T_{ex} <$ 10 K over
most of the G0.253+0.016 photosphere.
\citet{Clark2013}  modeled both the gas and dust temperatures in G0.253+0.016, 
finding that  high gas temperatures and lower dust temperatures can be explained by
a very high cosmic-ray flux and ionization rate.    The enhanced cosmic-ray flux
may also  enhance the abundances of certain molecules such as methanol
\citep{Yusef-Zadeh2013} and \HCOP .

\HCOP\   is the most abundant molecular ion next to H$_3^+$ in the
molecular phase  of the interstellar medium.   Collisional excitation of 
its  J=1-0 transition  requires  a collision partner density (H$_2$ 
and He) of  $n > 10^4$ cm$^{-3}$ to  thermalize. 
\HCOP\ has as relatively  high  abundance of
$n({\rm HCO^+}) / n({\rm H_2})  \sim   10^{-9}$ to $10^{-7}$ 
in nearby molecular clouds.     \HCOP\ is created by the reaction
$\rm{CO   + H_3^+ ~-> HCO^+   +   H_2}$ and destroyed by
reactions with H$_2$,  
$\rm {HCO^+  + H_2   ->  H_2CO  + H^+}$ or
by recombination with electrons,
$\rm{ HCO^+     + e^-   ->  CO  + H^+}$.   The large abundance of
$\rm { H_3^+}$ in the CMZ \citep{Goto2008, Goto2011}
may drive the \HCOP\ abundance in the CMZ to values near the 
upper-end of the above range.

The column densities of the absorption filaments can be estimated from the 
absorption optical depth.  Following \citet{LisztLucas2004}   the column 
density of a linear molecule having dipole moment $\mu$ (about 4 Debye 
for \HCOP ) absorbing from its ground  state is
$$
N_0 = {{8.0 \times 10^{12} \int \tau_{10} dV } \over { \mu^2 (1 - exp(-h \nu_{10} / k T_{ex})}}
$$
where $\nu_{10}$ is the J = 1$-$ 0 frequency (89.1885 GHz)  for \HCOP\ and 
$T_{ex}$ is the excitation temperature.
Measuring the \HCOP\  fluxed in the filament and comparing this with the 
surrounding off-filament pixels 
indicates that typically about one-half of the background light is absorbed, implying 
a typical absorption optical depth of about 0.7.
For the broad-line filaments,  $dV \sim$ 10 to 30 \kms\ and  $\tau_{10} \sim $0.2 to 2 
with a mean value of 0.7.
For a typical equivalent width (integral of absorption optical depth over the
line width) of 10 \kms , the   column density of \HCOP\  absorbers is
$N_0 \approx $ $10^{13}$ cm$^{-2}$ for 
$T_{ex}$ = 2.73 K, and $6 \times 10^{13}$ cm$^{-2}$ for  $T_{ex}$ = 10 K.

Table~2 lists the excitation temperatures ($T_{ex}$),  opacities ($\tau$), and
brightness temperatures  ($T_R$) for the J = 1$-$0 transition of \HCOP\  for a range of 
molecular hydrogen volume densities ($n$) and \HCOP\ column densities ($N$)
using  the radiative transfer code  RADEX   \citep{vanderTak2007}.   These 
calculations show that  an $H_2$ density of $1 \times 10^3$ cm$^{-3}$ an 
\HCOP\ column density of  $1 -  10 \times 10^{13}$ cm$^{-2}$ and line width 
$\Delta$V = 10 \kms\  in a gas with a kinetic temperature of 60 K results 
in an excitation temperature of about 2.8 $-$ 3.2 K, an optical depth 0.8 $-$ 7 in
the J = 1-0 line.   These parameters would produce the absorption seen in the
NLA filaments.  The BLA filaments require a slightly lower $H_2$ density to produce the
low observed brightness temperature and an opacity of order unity required to explain
the amount of background attenuation.   The \HCOP\ radiation field of 
G0.253+0.016 may contribute to the excitation of the upper state by effectively raising
the background radiation temperature (radiative pumping).  Detailed modeling of
the excitation conditions and radiative transfer using realistic cloud geometries is 
needed to determine if the low observed brightness temperature of the NLA and
BLA filaments can be used to place  a lower bound on separation between the 
absorbing gas and G0.253+0.016.   


The width of these filaments is less than about  2\arcsec\ or about 0.08 pc.  
(If only 1\arcsec\  wide, the filaments must be completely dark and only a lower
bound can be placed on their optical depth.  Higher angular resolution ALMA 
observations will be obtained during ALMA Cycle 1 to measure their actual widths.)
From the estimated column density and \HCOP\  abundance $X( HCO^+ ) = 10^{-7}$, 
assuming that the  depth of the filament is comparable to its projected width,  the higher 
excitation temperature (T$_{ex}$ = 10 K)  would  imply an H$_2$ column density of 
$6 \times 10^{20}$ cm$^{-2}$ and a volume density of around $2.5$ to 
$5  \times 10^3$~cm$^{-3}$.    A density larger than this value would drive the
absorption filaments into emission in \HCOP .   A lower \HCOP\ abundance increases 
the estimated volume density proportionally and conversely,  a higher
\HCOP\ abundance decreases the H$_2$  volume density.    Thus, either the  \HCOP\  
abundance must be higher than $X( HCO^+ ) = 10^{-7}$,  or the
line-of-sight depth of the filaments must be larger than their projected widths.    

It is possible that the filaments are sheets seen edge-on so that the 
line-of-sight depth is much larger than projected width of the filament.  In such a sheet, 
the column density to volume density ratio can be much larger than for a cylinder.   
For a given  \HCOP\ abundance,   the observed column density 
requires a volume  density that is  lower by the ratio of the  line-of-sight depth of the sheet  
divided  by its projected width on the plane of the sky.      Such edge-on sheets may be 
produced  by shocks propagating into molecular media at right angles to our line-of-sight.
With n(H$_2$) $< 5 \times 10^3$~cm$^{-3}$, the densities in the absorption 
filaments must be lower than the density at the $\tau \sim$ 1 surface in \HCOP\  
which is sub-thermally excited, and well below the mean density  of  G0.253+0.016,  
$n(H_2) \approx 7 \times 10^4$ cm$^{-3}$ 
\citep{Lis2001, Longmore2012,Rathborne2014_MALT90}.   

Inspection of the  \HCOP\ and HCN data  show that both tracers are heavily 
absorbed on the blueshifted, approaching side of the spectral line profiles.  Thus,
there is a pervasive layer of  sub-thermally excited, low density  \HCOP\ and HCN 
bearing gas in front of G0.253+0.016.     
If associated with G0.253+0.016, this gas would indicate the presence of 
an extended layer expanding from G0.253+0.016 at velocities 10 to  20 \kms\ over 
the entire face of this cloud.    

While NLA filaments near \Vlsr\ = 0 \kms\ could be located anywhere between the Sun
and  G0.253+0.016,  the higher radial velocity  NLA filaments with  \Vlsr\ $>$ 10 \kms\ are 
most likely to be in the CMZ (local gas is found between \Vlsr\   $\sim$ $-$10 to +10 \kms\
and the Galactic spiral arms between the Sun and the CMZ are  at  \Vlsr\   $\sim$ $-$
10 to $-$65 \kms ).  Near the projected edge of G0.253+0.016 some NLA filaments go 
faintly into emission  in  HCN and HNCO.  Because these
features disappear a few arc seconds beyond the cloud's edge,  they must be physically 
close to G0.253+0.016,  suggesting that they  trace  sub-thermal \HCOP\ and HCN molecules  
on the near surface of the cloud. 

\section{The Dynamic State of G0.253+0.016: Collapse, Expansion, or Both?}

\citet{Rathborne2014_MALT90} analyzed millimeter and sub-millimeter-wave emission in
multiple transitions from G0.253+0.016 obtained with the Mopra and APEX 
single-dish telescopes.   In addition to the large north-south velocity gradient 
they found that the optically thick dense-gas tracers were systematically redshifted 
with respect to the optically thin and hot gas tracers,  indicating radial motions which
could be interpreted as either due to expansion or collapse.   If the brighter, redshifted 
emission in optically thick species such as \HCOP\  and HCN originate in an 
externally-heated layer at the {\it front surface}, then the cloud is contracting.  
\citet{Rathborne2014_MALT90} refer to this as their `Baked Alaska' model.
The bright blueshifted emission expected from the far-side
heated layer would be absorbed by \HCOP\ molecules in the supersonically turbulent,  
colder, cloud interior.    This interpretation requires that the excitation temperature of the 
\HCOP\  J=1$-$0 transition in the cloud surface layer producing the redshifted line 
component must be higher than excitation temperature of the component producing 
the lower-velocity and lower brightness-temperature emission in the cloud interior.  
Alternatively, if the excitation temperature in the layer where the optical depths reach
a value of order unity is low compared to the excitation temperature deep in the cloud
interior, then these observations can be interpreted  as evidence for expansion of the 
cloud surface layers.    For sub-thermal excitation of molecules at the cloud surface,  
collapse would produce redshifted absorption while expansion would produce
blueshifted absorption.

At lower, blueshifted radial velocities, 
the \HCOP\  J=1$-$0 transition is seen as a network of absorption filaments against 
brighter background emission.   The brighter, background \HCOP\ emission appears to
fill the synthesized ALMA beam and to cover most of the projected surface area of
G0.253+0.016 with a peak brightness temperature of order 2 to at most 10 K,  well
below the gas kinetic temperature, $T_K > $ 60 K.   A peak line temperature lower 
than $T_K$ requires that either the emitting region fills only a small portion of the telescope 
beam (`beam dilution'), or a small fraction of the spectral channel (`frequency dilution'), 
or that the excitation temperature $T_{ex}$ at the  $\tau \approx$ 1 surface be much 
lower than $T_K$.    The bright \HCOP\ emission between the absorption filaments 
appears smooth on scales larger than the 1.7\arcsec\ ALMA beam, and continuous
from one velocity channel to the next,   indicating  that  beam and 
frequency dilution of the \HCOP\ emission is unlikely.  Thus, the relatively low brightness
temperature of \HCOP\ is most likely due to low excitation temperature, implying that
the H$_2$ density near the $\tau$ = 1 surface is much lower than the critical density 
for the \HCOP\ J = 1$-$0 transition.   

The \HCOP\  filaments are seen in absorption against this background 
emission,  they have an even lower excitation temperature than the regions seen
in emission.  If these blueshifted features are at the cloud surface,   the G0.253+0.016
surface layers must be expanding away from the cloud center of mass despite the large 
mass and gravity field.  This is consistent with the second (`P-Cygni') 
interpretation of the single-dish data  presented by  \citet{Rathborne2014_MALT90}.   

The mass-loss rate is difficult to estimate because the density on the absorbing 
layer is an upper bound, the column density is a lower-bound (based on an assumed
\HCOP\ abundance), and the thickness of the absorbing layer is not known. 
For a given density $n(H_2)$, and assuming that all sides of the cloud
are expanding in a manner similar to that side facing the Sun, the mass-loss rate 
is given by  $\dot M \sim  4  \pi f R^2_{eff} \mu m_H n(H_2) V$
where $f$ is the fraction of the cloud surface area occupied by 
expanding gas,  $ R_{eff} \approx 2.7$ pc is the  mean cloud radius, $n(H_2)$ 
is the area averaged  mean density of the NLA filaments at the 
cloud's surface, and $V \sim$ 10 to 20 \kms  is the expansion speed of the absorbing 
layer.   For $f $ = 0.5 and  $n(H_2) = 10^3$ cm$^{-3}$, set by the constraint that 
\HCOP\  in the absorbing  gas is severely  sub-thermally excited, 
$\dot M \sim $ 0.03 to 0.06 \msol\  yr$^{-1}$.   In the last 0.25 Myr,
since its close passage to Sgr A, G0.253+0.016 could have  lost 10 to 20\% of its mass. 

It is possible that  G0.253+0.016 is simultaneously undergoing {\it both} collapse and
expansion.   The presence of a maser and at least one massive clump surrounding it
suggests that at least one  moderate or high mass star is forming in the cloud interior. 
The cold and dense cloud interior may thus be experiencing gravitational collapse
and star formation.  However, the lack of internal heating and absence of any other
signposts of  young stars other than the maser source, suggests that the cloud is only 
in the earliest phases of star formation.   Blueshifted absorption in optically  thick 
species  such as  \HCOP\  over much of the cloud surface area  is an indication that 
outer layers may be expanding.

\section{Discussion}

The filamentary structure of molecular clouds has been recognized for decades 
\citep{BallyOrion1987}.   However, the extensive arc-second scale \HCOP\  absorption 
filaments revealed by  ALMA are new.   The broad-line (BLA) filaments in G0.253+0.016 are 
extraordinary because of their narrow minor axis dimension (less than $\sim 10^4$ AU),  
their  20 \kms\ width in absorption, and their low antenna temperatures.    
The more abundant  narrow-line  (NLA) 
filaments are less  extreme in their velocity extent.   Comparison of the optically 
thick \HCOP\ and  HCN cubes with the other ALMA spectral line
cubes of optically thinner tracers such as HNCO, SiO, C$_2$H,  H$_2$CS,  
\hcn , and \hcop\  show that \HCOP\ and HCN are heavily
absorbed between V$_{LSR}$ = $-2.8 $ and about $+30$  \kms.    The BLAs
are only seen in \HCOP\ and have no corresponding features in any other species in 
either emission or absorption. 

Two models are considered for the BLA filaments.  The BLAs may be 
associated with dense  gas in the Galactic Center Ring (GCR) a portion of which  
has a radial velocity \Vlsr\ = 10 to 30 \kms\ towards Galactic longitude of 0.18  
\citep{RodriguezFernandez2006, RodriguezFernandez2008}. 
Assuming that the GCR  has similar orientation and kinematics but is larger 
than the  100 pc radius ring proposed by \citet{Molinari2011}, the front side 
is expected to have positive  radial velocities while the far side is expected to 
have negative velocities.   Given the  possible association of G0.253+0.016 
with the Molinari ring, the front-side of  GCR may be in front of 
G0.253+0.016.  Low density, low-excitation gas could then be seen
in absorption against the warmer emission from the cloud.   In this scenario 
the BLAs are produced by sub-thermal \HCOP\  far in front (a few to as much 
as $\sim$200 pc in front of G0.253+0.016).    Their narrow widths,  highly 
supersonic velocity extents, and elongated filamentary morphology  suggest 
that  they are ram pressure confined.  In this model,  they may be caustic surfaces 
produced where shock fronts are moving in the plane of the sky and whose 
post-shock layers are sheets observed 
very close to edge-on.   Thus, their depth along the line-of-sight may be much 
larger than their observed widths.   The \HCOP\ molecules must  either survive 
passage through the shock,  or are rapidly formed in the post-shock cooling layers.    
The crossing time of these filaments is about $10^3$ years.    If molecule reformation
is  responsible for the \HCOP\ abundance, then this time-scale  establishes a 
limit on the formation time.

The alternative model is that the BLAs are related to the non-thermal filaments 
near Galactic  longitude 0.18 and are magnetically confined.  This picture is 
motivated by the similar orientations of the BLAs and the 20-cm 
non-thermal filaments that lie only a few parsecs in projection from the 
southeast edge of G0.253+0.016.   Both sets of filaments are oriented roughly at
right angles to the Galactic plane.   Using equipartition arguments applied to the
synchrotron, 
\citet{YusefZadehMorris1987} estimated  that the magnetic fields in the 
non-thermal streamers have strengths of about  0.05 to 0.2 milli-Gauss.  

The polarization of the dust continuum emission from G0.253+0.016 has been 
measured at $\lambda$ = 350 $\mu$m \citep{Dotson2010}.   The magnetic field 
is nominally orthogonal to the 350 $\mu$m polarization vector.  Inspection of 
Figure 39 in \citet{Dotson2010}  shows that the magnetic field in the cloud interior 
where most of the sub-mm dust emission originates,  closely follows the bow-shaped 
emission-line filaments observed in the optically-thin tracers.   The magnetic field 
in the northern part of G0.253+0.016 is approximately orthogonal to the Galactic 
plane and parallel to the BLA filaments.  The magnetic field in the southern part 
of G0.253+0.016 is roughly aligned with the Galactic plane.    

Application of the Chandrasekhar-Fermi  method to estimating 
magnetic field strengths in the molecular gas in the CMZ as traced by 
sub-millimeter  polarimetry of the dust continuum emission leads to
field strength estimates of  1 - 3  milli-Gauss \citep{Chuss2003}.    
Using the scaling relations of magnetic field strength with cloud density 
based on Zeeman effect measurements \citep{Crutcher2012} and assuming
a mean cloud density of few~$\times 10^5$~cm$^{-3}$ leads to similar field strength
estimates.   \citet{Ferriere2010} presented a review of CMZ  magnetic fields.

An ionization fraction of  $\sim 10^{-7}$ (the estimated abundance of \HCOP\
in the absorbing medium) implies  very long  ambipolar diffusion time-scales and
highly elongated structures if magnetic fields are present in the absorption region.   
Heavy molecular ions such as \HCOP\    gyrate around the magnetic field lines 
at a frequency   $f_{gyro} \sim e B / m(HCO^+) c \approx 0.3 ~ B_{mG}$ 
radians per second.   Here,  $e$ is the electron charge, $c$ is the speed of light,
$m_{HCO+} = 29 ~ m_H$ 
is the mass of an \HCOP\ molecule,  and  $B_{mG}$ is the magnetic field in units of 
1 milli-Gauss.   Using the observed spectral line half-width of the most prominent 
BLA filament,  $\Delta V \sim$10 \kms , implies a gyro-radius of \HCOP\ molecules 
of $\sim$ 30 km.  The  mean-free path of molecules,   
$\lambda _{mfp} \sim 1 / n(H_2) \sigma$  
for  a density $n(H_2) = 10^{3}$ and a collision cross-section 
$\sigma \sim 10^{-15}$~cm$^{-2}$ is $\lambda _{mfp} \sim 10^{12}$ cm,   5-orders 
of magnitude smaller than the width of the BLA filaments ($\sim 10^{17}$ cm).  
The mean time between molecular collisions which lead to diffusion 
is $ t_{col} \sim 10^{6}$ seconds.   To random-walk a distance $10^5$ times larger
than the mean free-path requires about $10^{10}$ collisions.  Thus, the time for 
molecules to diffuse over a distance of $r_{filament} \sim10^4$ AU (the typical width of the 
absorption filaments) is therefore $t_{AD} \sim10^{10} t_{col} \sim 10^{16}$ 
seconds.  For the parameters implied by the observations,  the magnetic field 
and the partially ionized molecular gas are tightly coupled.     The effective diffusion 
velocity {\it across} the filaments is 
$V_{perp} \sim r_{filament} / t_{AD} \sim$ 10 cm~s$^{-1}$.  
However, atoms and  molecules can move unimpeded {\it along}  the field lines 
at the $\Delta V \sim 10$ \kms\   measured  velocity dispersion.    Thus the aspect 
ratio (length-to-width) of magnetized filaments can be as high as  
$\Delta V  / V_{perp} > 10^5$. 
 
\subsection{Expansion of the Surface G0.253+0.016: Photo-ablation?}

Two mechanisms could be responsible for the apparent expansion of the near-side surface layers 
of G0.253+0.016;   photo-ablation and stored magnetic energy. 
The first  mechanism relies on  external heating of the cloud surface layers by the combined
effects of UV, soft X-ray, visual, and near-IR radiation that can photo-ablate
cloud surfaces, generate large internal motions, and generate shocks which compress the cloud
surface layers \citep{Mellema2006,Henney2009M,Arthur2011}.   Diffuse dust emitting between
8 and 70 $\mu$m tends to be associated with photo-ionized plasmas in HII regions.
The extended 20 cm free-free  continuum \citep{YusefZadeh2004}  and 8 to 70 $\mu$m 
continuum emitted by diffuse mid-IR emission from warm  dust implies that much of the 
inner 100 pc of the Galaxy contains photo-ionized plasma and 
warm dust, likely produced by  radiation from
massive stars.  Although the Spitzer and WISE mid-IR images show that the Galactic Center
Bubble located between Sgr A and G0.253+0.016, and likely ionized and heated by the Arches
and Quintuplet clusters,  dominates the diffuse mid-IR emission in the CMZ,  these images 
also exhibit somewhat fainter and more diffuse warm dust emission centered about 
6\arcmin\  (15 pc in  projection) due East of G0.253+0.016 at 
$[l,b]$ = $[0.304, -0.067]$ (J2000 =  17:46:36.2, -28:42:41).   

Diffuse 20 cm radio continuum and several HII regions within 6\arcmin\  of this region of 
peak infrared emission (but generally due East of G0.253+0.016) 
indicate the presence of OB stars whose energy output may irradiate and heat the G0.253+0.01 
cloud's surface.  If the Lyman continuum luminosity of these stars is $Q_{50} \sim 10^{50}$ 
ionizing photons per second, and located at a distance $D_{15}$ = 15 pc, the electron 
density  at the surface of G0.253+0.016 would be (neglecting any intervening extinction)  
about $n_e = (Q / 4 \pi \alpha _B r)^{1/2} D^{-1}$
$\sim 40~Q_{50}^{1/2} r_3^{-1/2} D^{-1}_{15}$ cm$^{-3}$
assuming a spherical cloud of radius $r  = r_3 \sim $ 3 pc 
(here, $\alpha _B = 2.6 \times 10^{-13}$ cm$^3$~s$^{-1}$ is the Case B hydrogen
recombination coefficient).     But, G0.253+0.016 appears filamentary with local radii of 
curvature at least  an order of magnitude
smaller in at least  one direction.   Thus, the actual electron density may be $n_e >  10^2$ cm$^{-3}$. 
The implied mass-loss rate due to heating (assuming that the gas is accelerated to about
the sound-speed in ionized gas;  $\sim$ 10 \kms )  is about $10^{-3}$ \msol\ yr$^{-1}$ or 
only about $10^2$ to $10^3$  \msol\  in 0.25 Myr.   This is an order-of-magnitude lower than
the mass-loss rate estimated from the NLA absorption filaments.  Unless additional energy
input mechanisms operate, photo-ablation is an unlikely source of cloud surface layer 
expansion.

\subsection{Expansion of the Surface G0.253+0.016: Magnetic Bounce?}

 \citet{Longmore2013} proposed 
that G0.253+0.016 passed near Sgr A about 0.25 Myr ago.    Passage of the cloud through the 
deepest part of the Galactic gravitational potential would have compressed the cloud.    
If its interior magnetic fields were sufficiently tangled, they would act as a spring, 
storing the energy  of compression,  limiting the dissipation of internal motions by shocks.   
As G0.253+0.016 emerged from the 
potential well,   the compressed fields could drive re-expansion of the cloud.   While some regions
with high density or mass-to-flux ratios may collapse, less dense or relatively weakly bound
parts of the cloud such as its outer layers, may expand as the fields relax.
  
An order of magnitude estimate of the mean field strength required to support such
a `bounce'   can  be obtained by assuming that the stored magnetic energy, 
$E_B \sim f B^2 r^3$ is comparable  to the gravitational self-energy $E_G \sim g G M^2 / r$.  
Here,  $f$ and $g$ are  factors of order  unity depending on the magnetic field geometry, 
the cloud shape,  and density  distribution, and for G0.253+0.016, $r \sim$ 3 pc  is the mean 
cloud radius,  and $M \sim 10^5$ \msol\  is the cloud mass.  If the gravitational potential 
energy is similar to the magnetic energy (as required by a `bounce'), these parameters imply a 
mean magnetic  field strength $B \sim $ 1.5 milli Gauss, consistent with the magnetic
field-strength estimates.    Thus, the adiabatic behavior of a 
tangled internal magnetic  field could drive a re-expansion of G0.253+0.016 following its emergence 
form the deepest  part of the Galactic gravitational potential.

Future work should  search for evidence for BLAs in other optically-thick ground-state transitions  
of ions  such as N$_2$H$^+$.   BLAs detected only in  other ions would lend 
support to the magnetic structure interpretation.  BLAs detected in other neutral species, especially
ones with enhanced abundances in shocks, would provide support for the shock-wave
interpretation.  High resolution dust continuum polarization measurements and searches for 
Zeeman splitting are needed to constrain the magnetic field geometry and strength to determine if
the magnetic flux-to-mass ratio is super-critical and allows the cloud to collapse, or sub-critical
and may prevent collapse and formation of a cluster of stars.

\section{Conclusions}

ALMA Cycle 0 observations of the massive, compact, and dense Galactic center cloud, 
G0.253+0.016 are presented with an angular resolution of about 
1.7\arcsec\ in the \HCOP\  and \hcop\  J = 1$-$0 lines.  The main results derived from 
these observations are:

[1]  The observed intensity ratios of \HCOP\ and \hcop\ indicate that the mean 
optical depth of \HCOP\ is around unity or higher. 
The observed  J=1$-$0 \HCOP\ brightness temperature
of 2 to 12K  in a cloud in which the kinetic temperature of the gas is estimated
to be around 60 to 80 K indicates sub-thermal excitation at the \HCOP\ photosphere and
an H$_2$ density (n(H$_2$)~$\sim 10^3$~cm$^{-3}$) more than an order of magnitude
lower than the critical density for the J=1$-$0 transition and more than an 
order of magnitude lower than the mean density of G0.253+0.016. 

[2]  The ALMA observations reveal a network of  blueshifted (relative to the
cloud)  absorption filaments seen against the sub-thermally excited \HCOP\ emission.
These filaments must have densities even lower than density at the $\tau_{12} \sim$ 1
photosphere where most of the \HCOP\ emission is produced.
A few  broad-line absorption (BLA) filaments have line widths of over 20 \kms ,  
widths of only a few arc seconds,  and lengths of nearly 1 arc minute.
They are only seen in \HCOP .    The BLA filaments may be foreground magnetic structures 
possibly associated with the brightest  non-thermal filaments that cross the Galactic plane at 
longitude 0.18\arcdeg .  Alternatively, they may be shock fronts in foreground CMZ clouds 
propagating orthogonal to the line-of-sight so that their compressed layers are seen
nearly edge-on with a depth along the line-of-sight much greater than their projected widths.
Additionally, G0.253+0.016 is laced with dozens of narrow-line absorption (NLA) filaments with
line widths of less than  20 \kms .    Most of these features are seen on the blueshifted
side of the line profiles.  A few have counterparts in HCN.    
Some absorption filaments seen in both \HCOP\ 
and HCN which extend over faint HCN backgrounds go into emission in HCN but not in
\HCOP  .   HCN also shows extensive absorption
on the blueshifted side of G0.253+0.016.   These characteristics suggest that the NLA filaments
trace sub-thermally excited, optically thick gas on the front surface of G0.253+0.016.  

[3] The blueshift of the absorption filaments relative to the peak \HCOP\ 
emission indicates that the surface layers of G0.253+0.016 are 
expanding with velocities larger than 5 to 20 \kms\  with respect to the cloud's center of mass.  
The apparent ``P-Cygni" line profile  may be an indication of  re-expansion of the 
G0.253+0.016 surface layers  following a  recent  passage close to Sgr A.  
Compression of a milli-Gauss magnetic field  during the passage, followed by subsequent 
expansion driven by stored magnetic energy may be responsible.    

[4]  Weather or not G0.253+0.016  will form a star cluster or only a few stars remains unclear. 
Although the \HCOP\ absorption indicates that the cloud surface layers are expanding, the 
cloud interior may be in the earliest stages of star formation as indicated by the presence of at
least one  cloud core with an associated water maser.   The dynamic state of the dozens of
other 3 mm dust condensations in G0.253+0.016 remains unclear.    Measurements of the
amplitude and geometry of the magnetic field are needed to determine if the magnetic pressure
is sufficient or insufficient to prevent this massive, high-density, and compact cloud from 
collapsing to form a cluster or association of stars.

\acknowledgments{This research of JB and AG was in part supported by  National Science 
Foundation (NSF) grant  AST-1009847.   Observations reported here were obtained at the 
Atacama Large Millimeter Array (ALMA) 
located in Chile.   The authors acknowledge our ALMA Contact Scientist, Dr Crystal Brogan, 
for assistance in preparing the observations and performing the initial
data calibration.  J.M.R acknowledges funding support via CSIRO's Julius Career Award 
that was used to host a week-long team meeting which laid the foundation for this work. 
J.M.R, S.N.L, J.M.D.K, J.B. and N.B. acknowledge the hospitality of the Aspen Center for Physics, 
which is supported by the National Science Foundation Grant No.~PHY-1066293. 
This paper makes use of the following ALMA data:  ADS/JAO.ALMA\#2011.0.00217.S. 
ALMA is a partnership of the European Southern Observatory (ESO) 
representing member states, Associated Universities Incorporated (AUI) and the National
radio Astronomy Observatories (NRAO) for the National Science Foundation (NSF) in the USA,  
NINS  in Japan, NRC in Canada, 
and NSC and ASIAA in Taiwan, in cooperation with the Republic of Chile.  The Joint ALMA
Observatory (JAO) is operated by ESO (Europe) , AUI/NRAO  (USA), and NAOJ (Japan).
The National Radio Astronomy Observatory is a facility of the National Science Foundation 
operated under cooperative agreement by Associated Universities, Inc.
}

\clearpage

\bibliography{Brick_filaments}

\clearpage

\input{Figures}

\input{Figures2}

\clearpage

\input{Table1}

\input{Table2}

\end{document}

%% file: Figures.tex
\begin{figure}
\epsscale{1.0}
\center{\includegraphics[width=0.8\textwidth,angle=0]
   {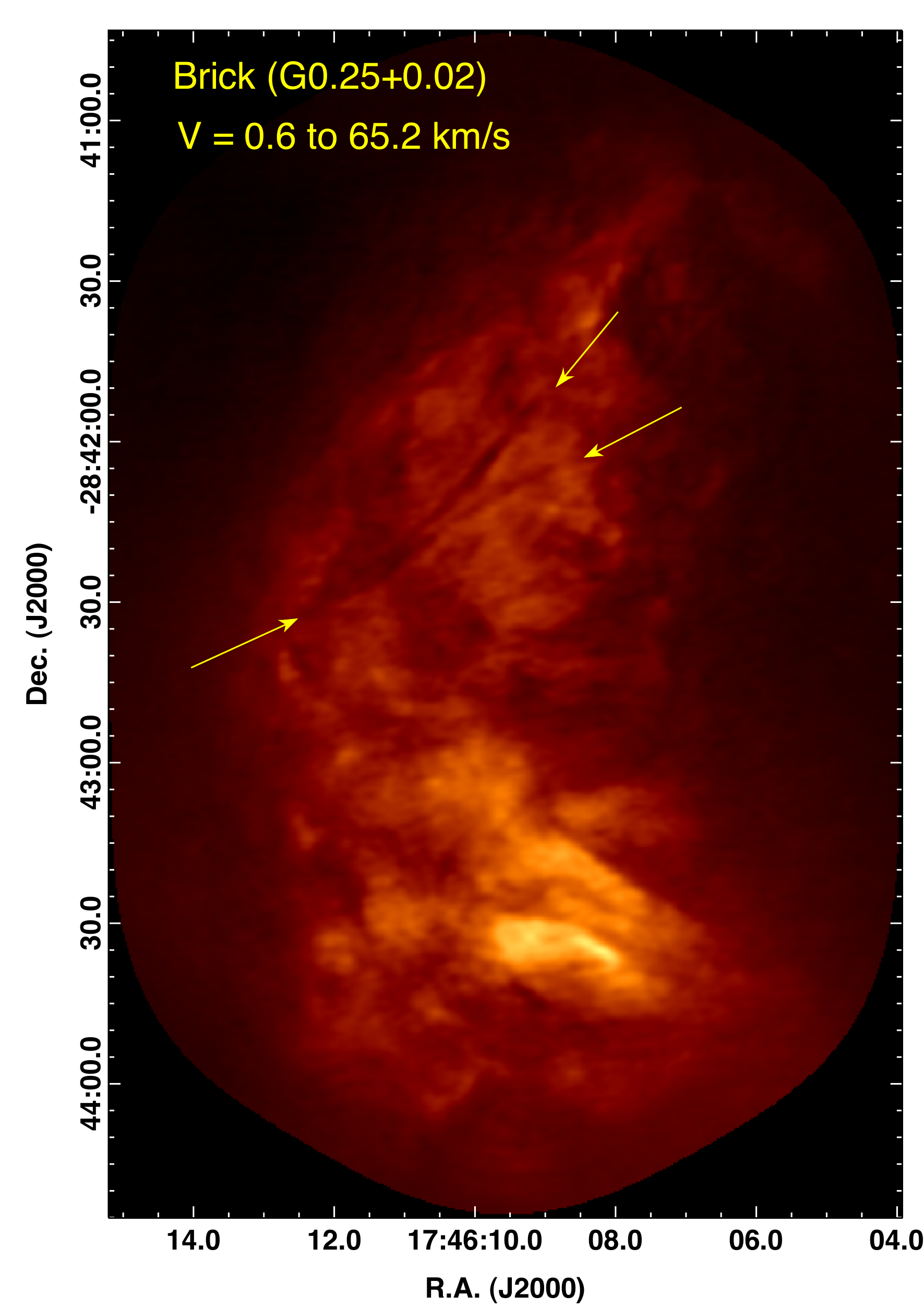}}
\caption{A velocity integrated image shoving G0.253+0.016
in the HCO$^+$ J = 1$-$0 transition. 
The velocity range extends from \Vlsr\ = 0.6 \kms\ to 65.2 \kms .  
Arrows mark the locations of the two broad line absorption (BLA) filaments.  
These  arrows are retained in Figures 2 through 9.}
\label{fig1}
\end{figure}

\begin{figure}
\epsscale{1.0}
\center{\includegraphics[width=0.8\textwidth,angle=0]
   {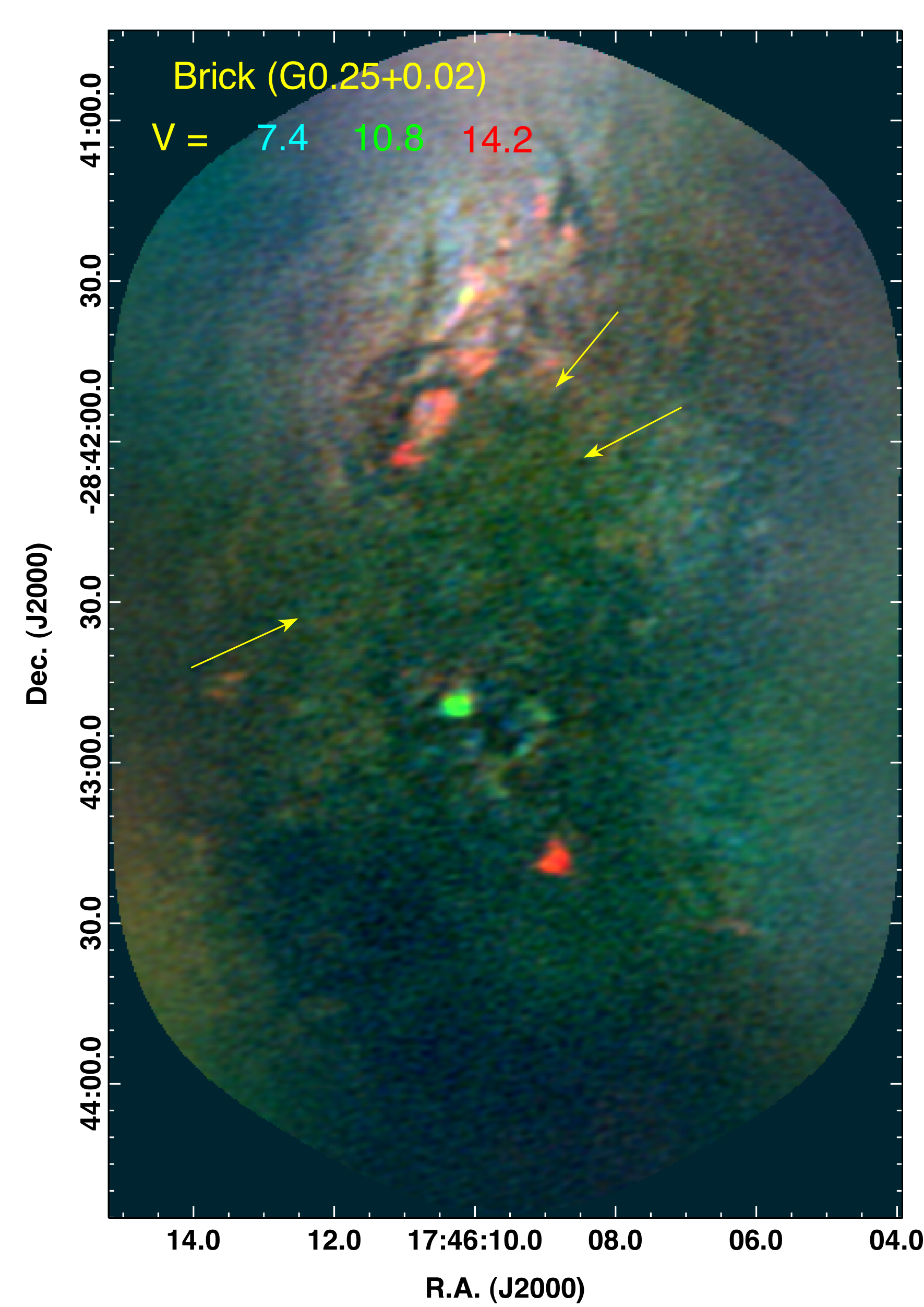}}
\caption{ An \HCOP\ image of G0.253+0.016  showing 3.4 \kms\ wide channels 
centered  at   $\rm V_{LSR} $ = 7.4 (blue), 10.8 (green), and 14.2 (red) \kms . 
In addition to the arrows marking the location of the most BLA filaments, the
three arrows in the lower-right mark the general locations
of narrow line absorption (NLA) filaments in Figures 2 through
9.
}
\label{fig2}
\end{figure}

\begin{figure}
\epsscale{1.0}
\center{\includegraphics[width=0.8\textwidth,angle=0]
   {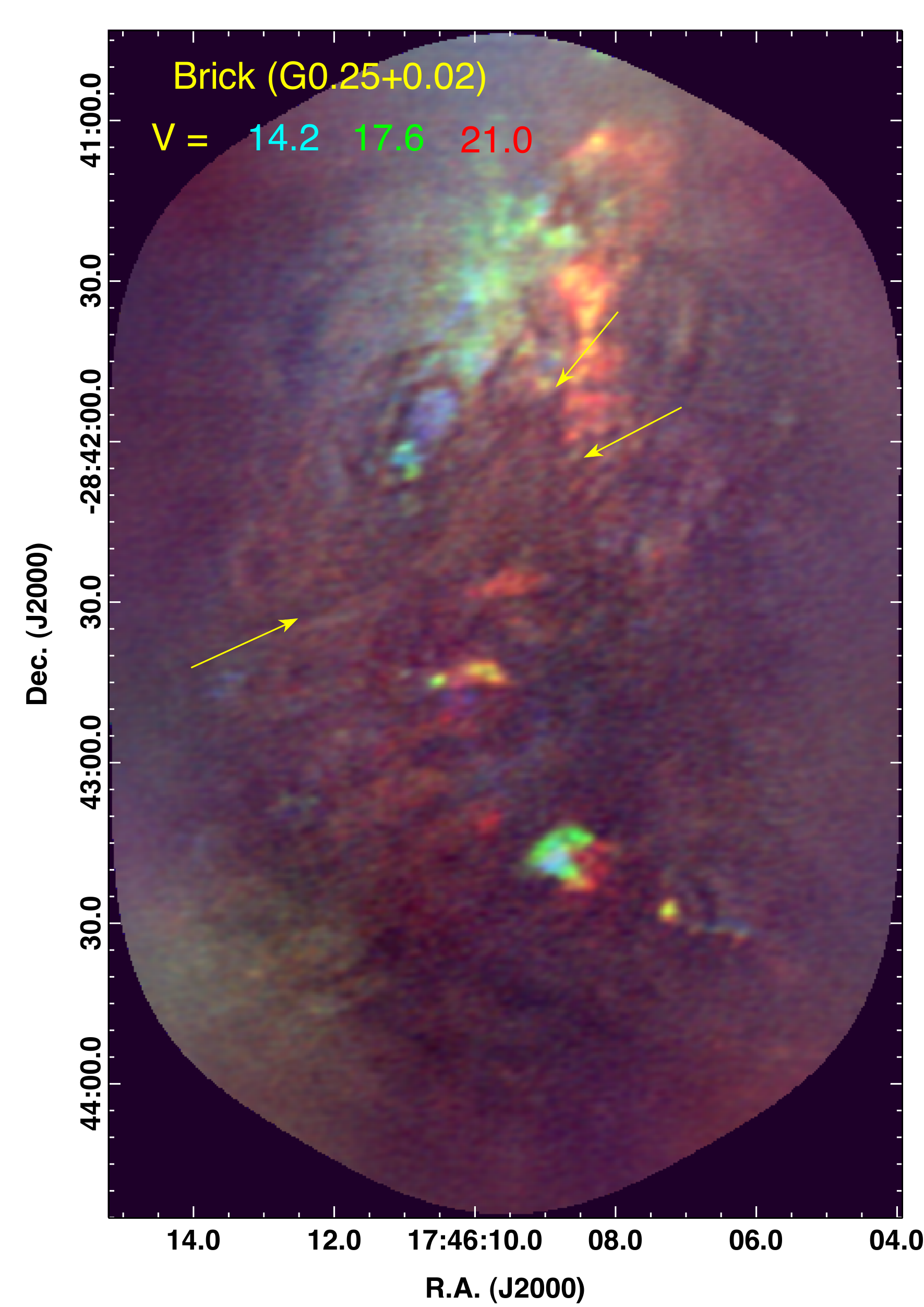}}
\caption{
This figure shows \HCOP\ channels centered at 
$\rm V_{LSR} $ = 14.2 (blue), 17.6 (green), and 21.0 (red) \kms . }
\label{fig3}
\end{figure}

\begin{figure}
\epsscale{1.0}
\center{\includegraphics[width=0.8\textwidth,angle=0]
   {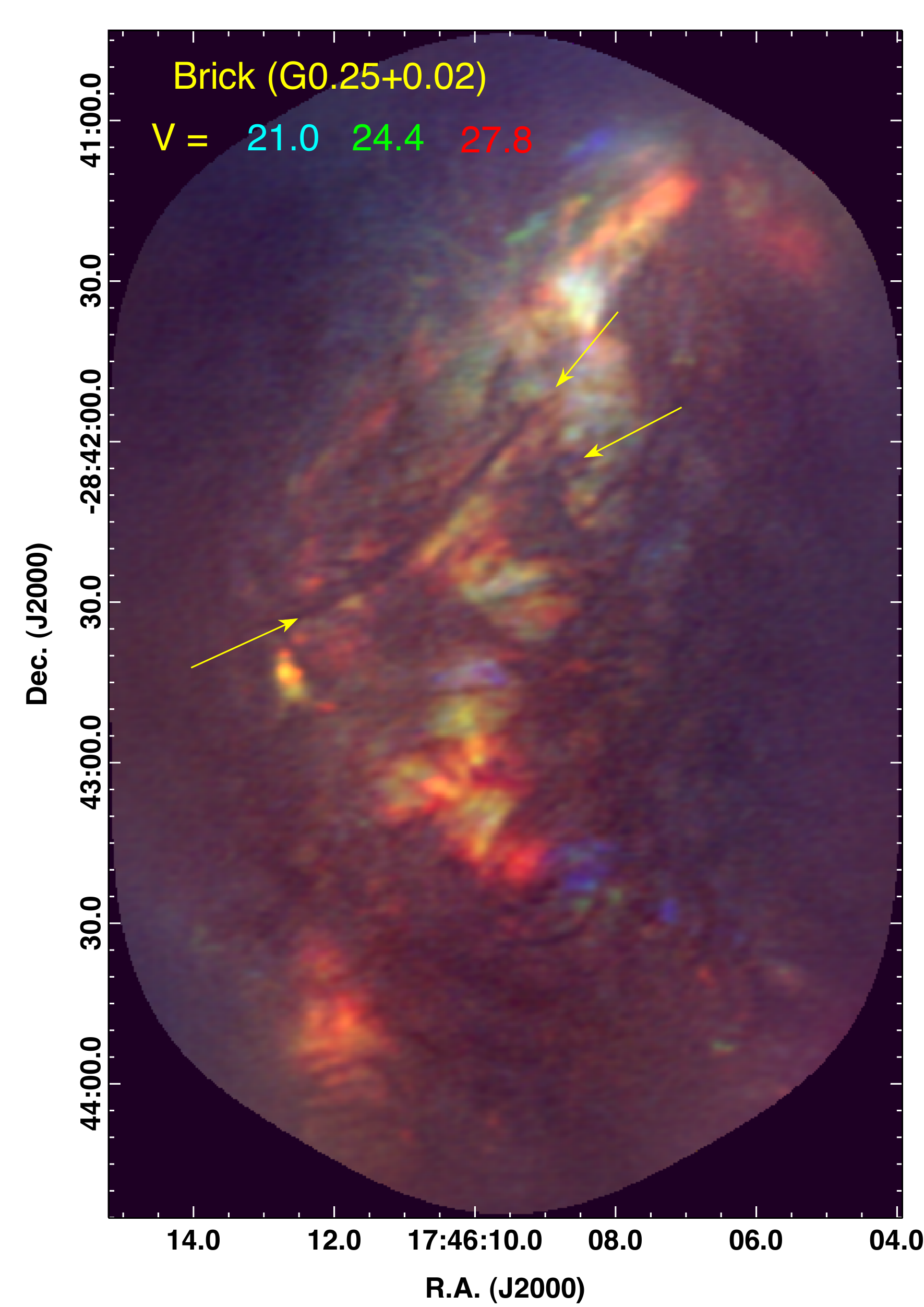}}
\caption{
This figure shows \HCOP\ channels centered at 
$\rm V_{LSR} $ = 21.0 (blue), 24.4 (green), and 27.8 (red) \kms . }
\label{fig4}
\end{figure}

\begin{figure}
\epsscale{1.0}
\center{\includegraphics[width=0.8\textwidth,angle=0]
   {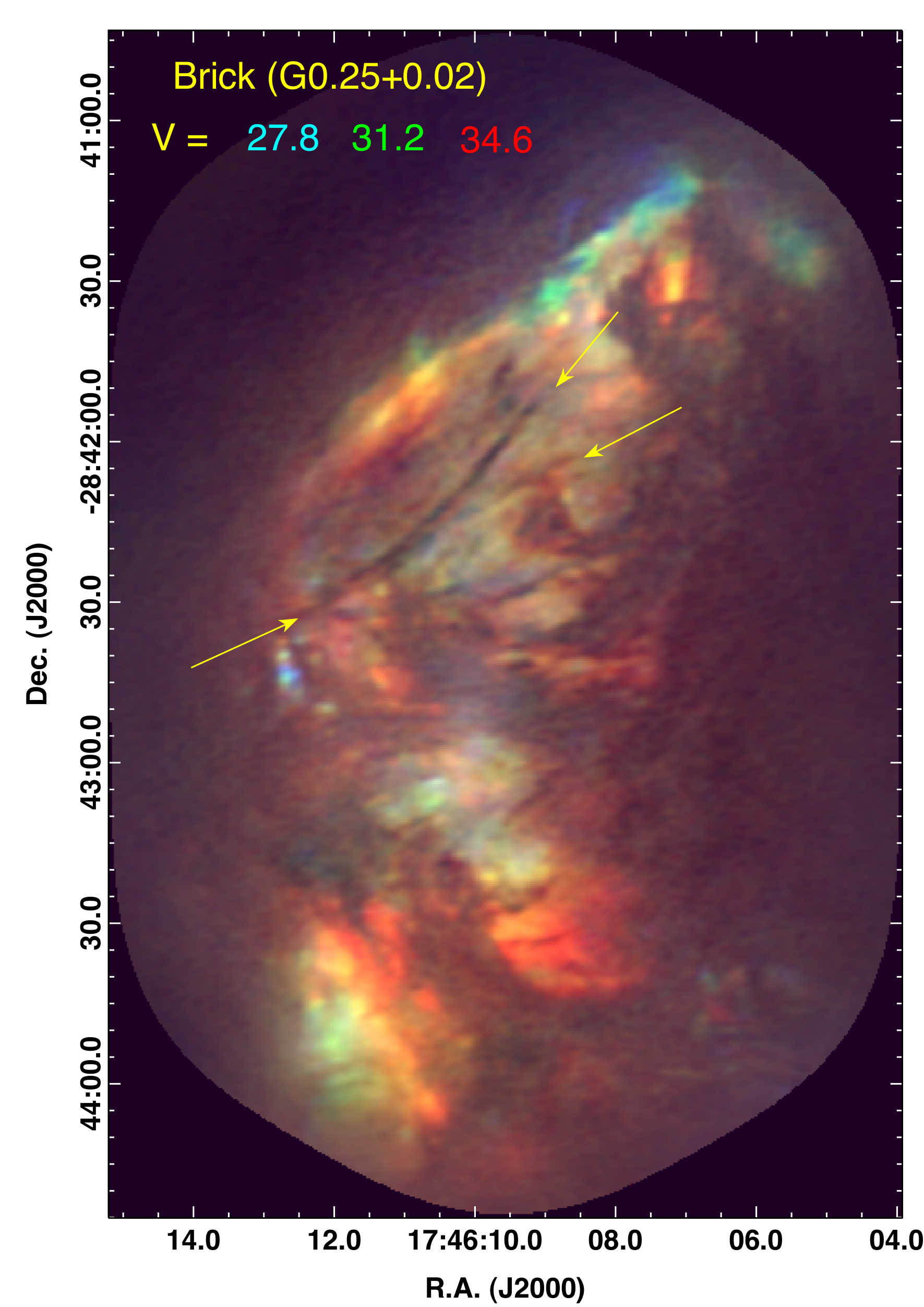}}
\caption{
This figure shows \HCOP\ channels centered at 
$\rm V_{LSR} $ = 27.8 (blue), 31.2 (green), and 34.6 (red) \kms . }
\label{fig5}
\end{figure}

\begin{figure}
\epsscale{1.0}
\center{\includegraphics[width=0.8\textwidth,angle=0]
   {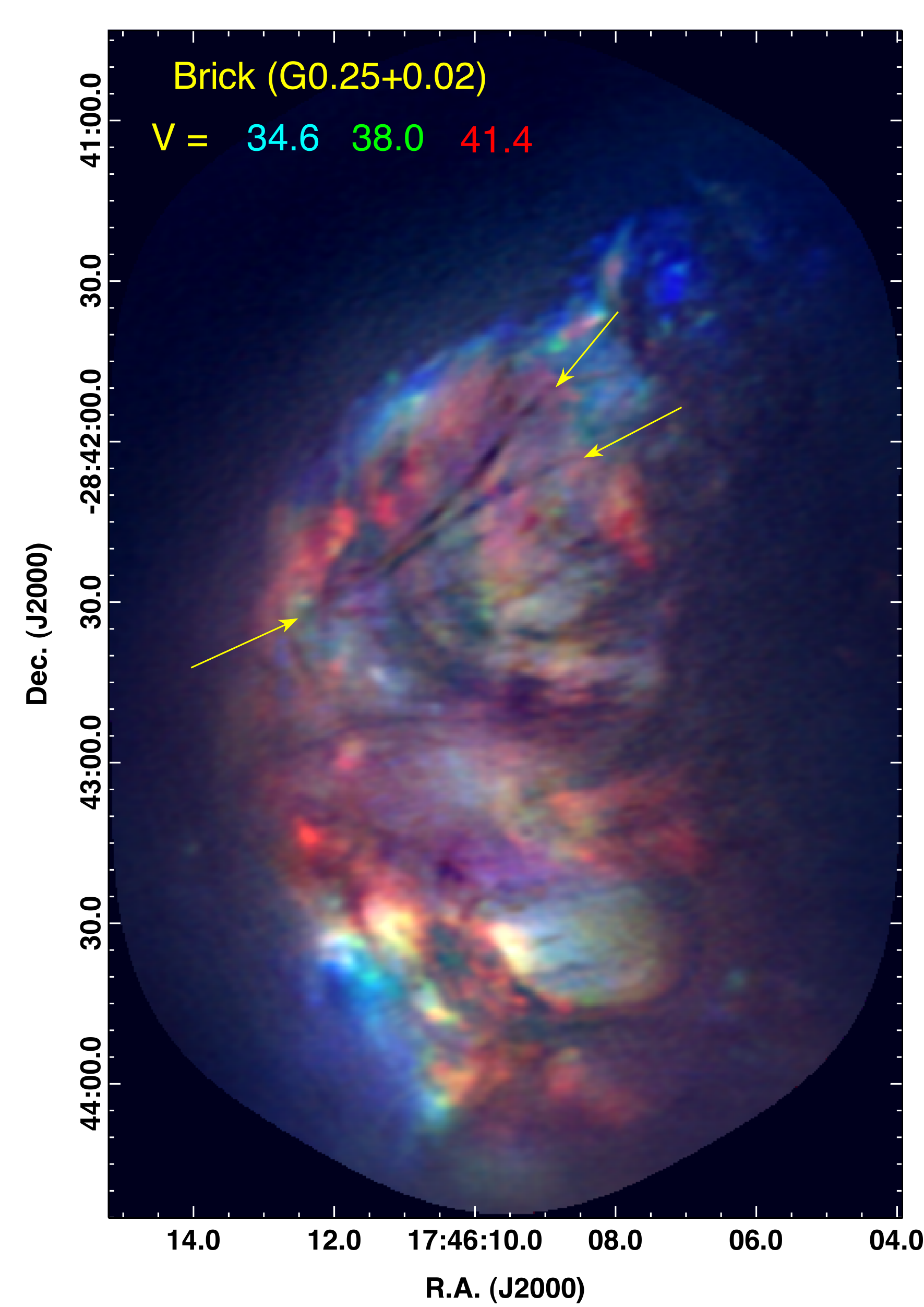}}
\caption{
This figure shows \HCOP\ channels centered at 
$\rm V_{LSR} $ = 34.6 (blue), 38.0 (green), and 41.4( red) \kms . 
}
\label{fig6}
\end{figure}

\begin{figure}
\epsscale{1.0}
\center{\includegraphics[width=0.8\textwidth,angle=0]
   {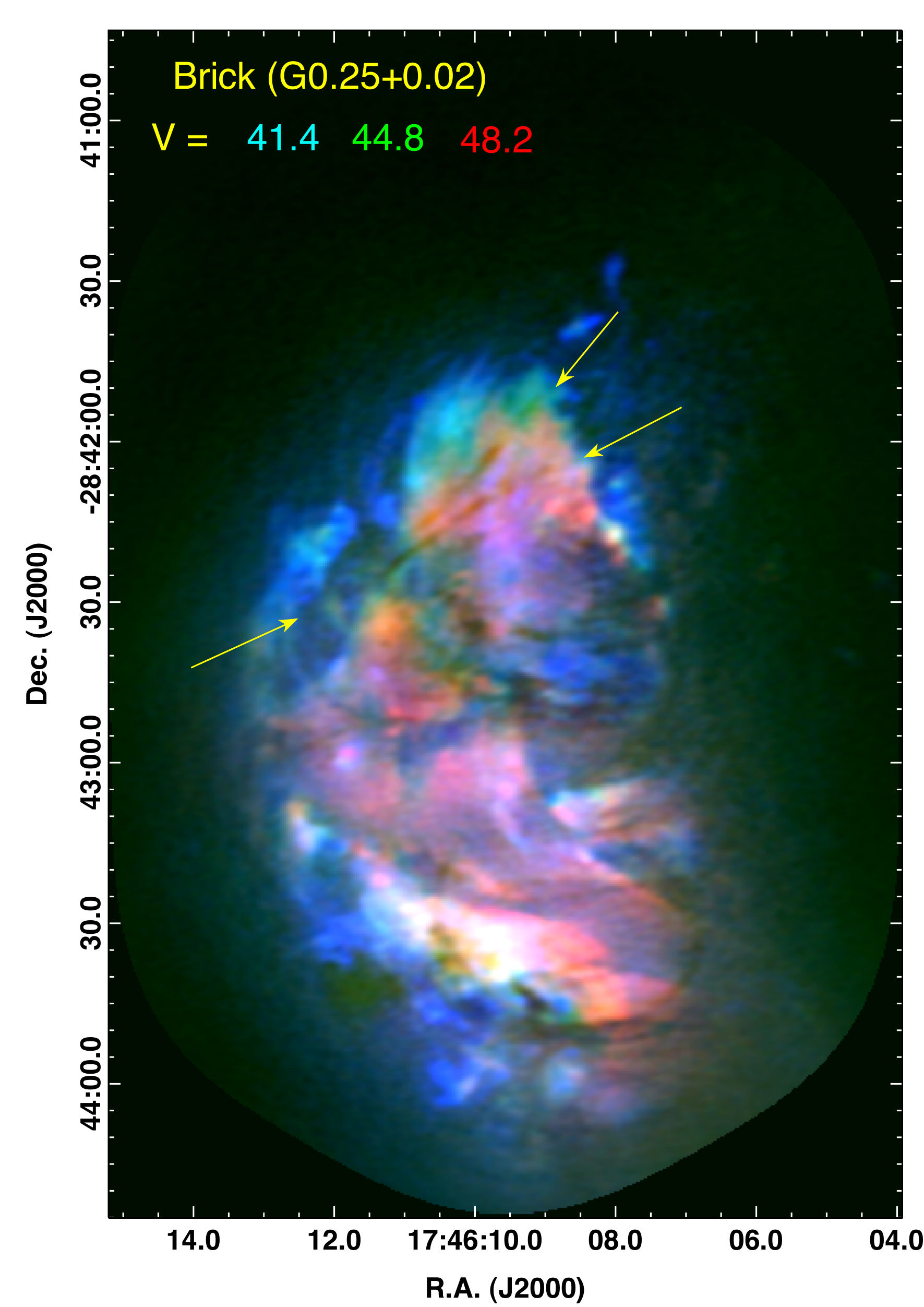}}
\caption{
This figure shows \HCOP\ channels centered at 
$\rm V_{LSR} $ = 41.4 (blue), 44.8 (green), and 48.2 (red) \kms . 
}
\label{fig7}
\end{figure}

\begin{figure}
\epsscale{1.0}
\center{\includegraphics[width=0.8\textwidth,angle=0]
   {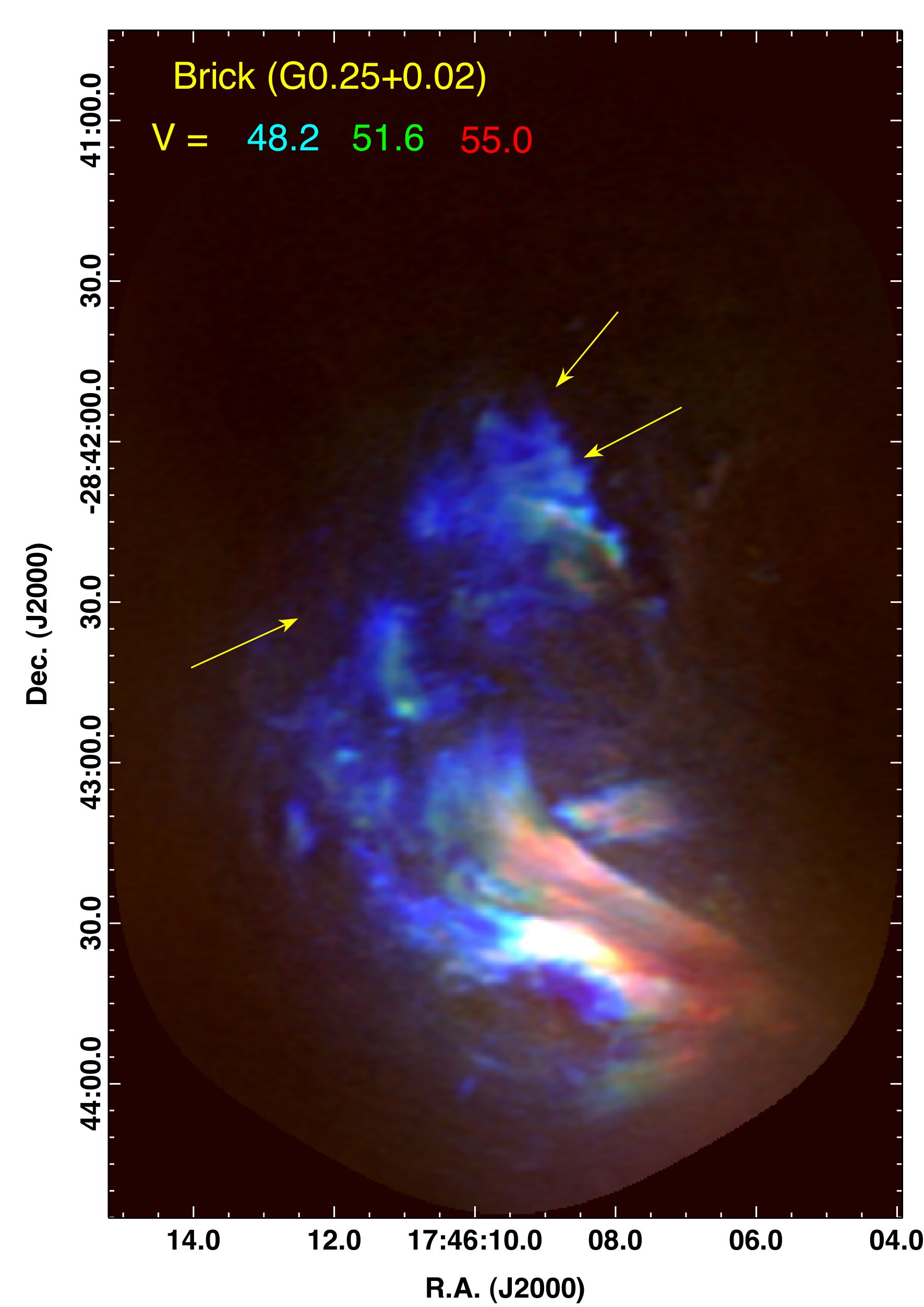}}
\caption{
This figure shows \HCOP\ channels centered at 
$\rm V_{LSR} $ = 48.2 (blue), 51.6 (green), and 55.0 (red) \kms . 
}
\label{fig8}
\end{figure}

\begin{figure}
\epsscale{1.0}
\center{\includegraphics[width=0.8\textwidth,angle=0]
   {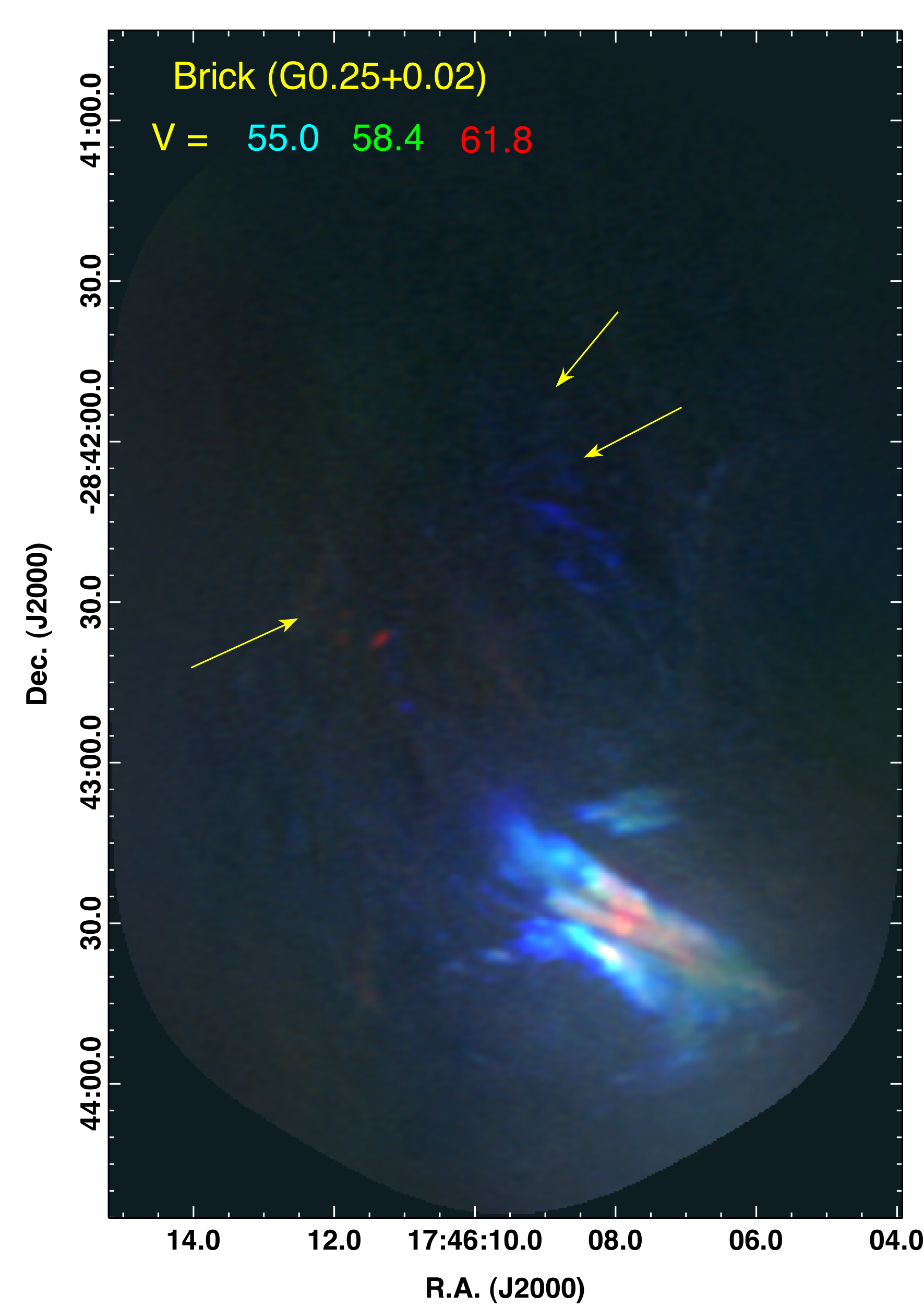}}
\caption{
This figure shows \HCOP\ channels centered at 
$\rm V_{LSR} $ = 55.0 (blue), 58.4 (green), and 61.8(red) \kms . 
}
\label{fig9}
\end{figure}

%% file: Figures2.tex
\begin{figure}
\epsscale{1.0}
\center{\includegraphics[width=1.0\textwidth,angle=0]
         {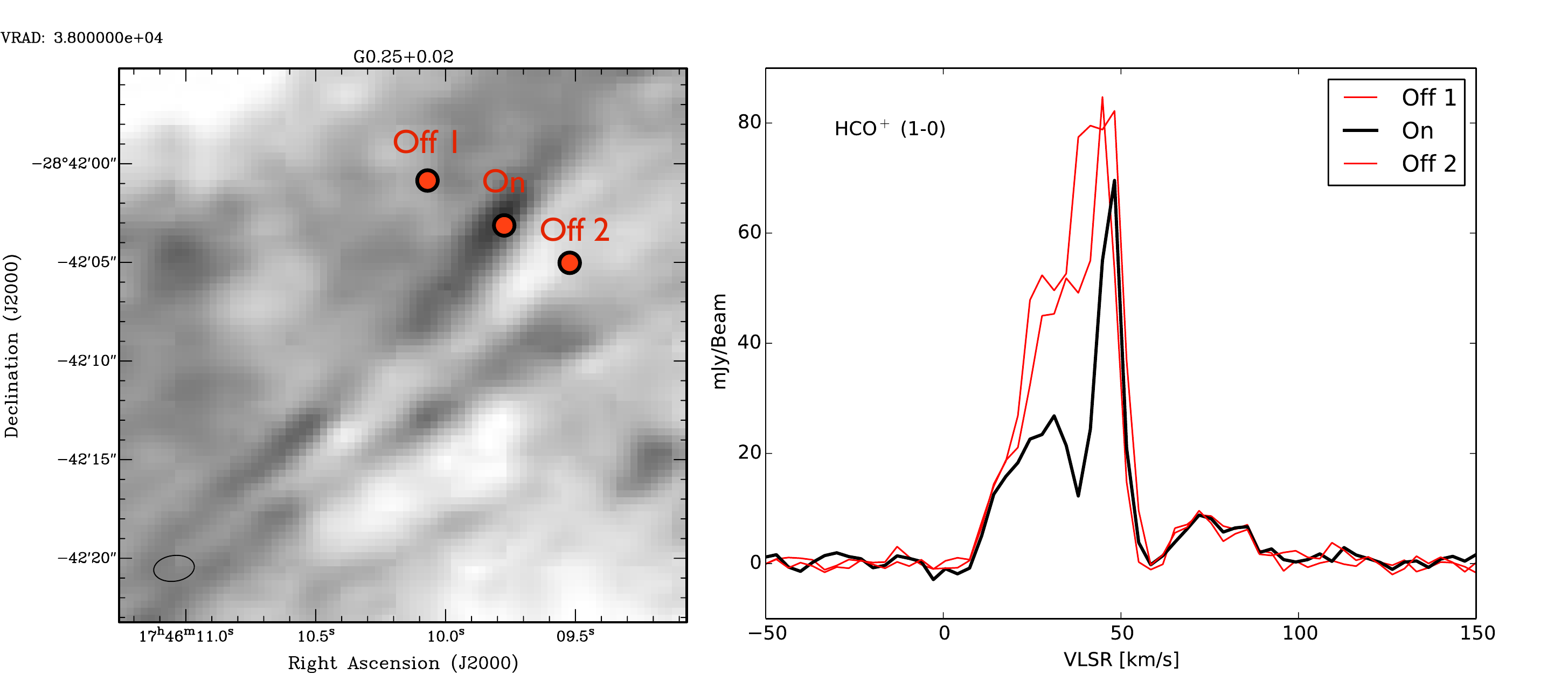}
         }
\caption{ {\it(Left:)} The location of three \HCOP\ spectra shown in the Right panel
                 {\it(Right:)}  The \HCOP\ spectra shows the most prominent BLA absorption 
                 at the location labeled  
                 `On'  (thick black line) and two off positions (thin red lines).
}
\label{fig10}
\end{figure}

\begin{figure}
\epsscale{1.0}
\center{\includegraphics[width=1.0\textwidth,angle=0]
         {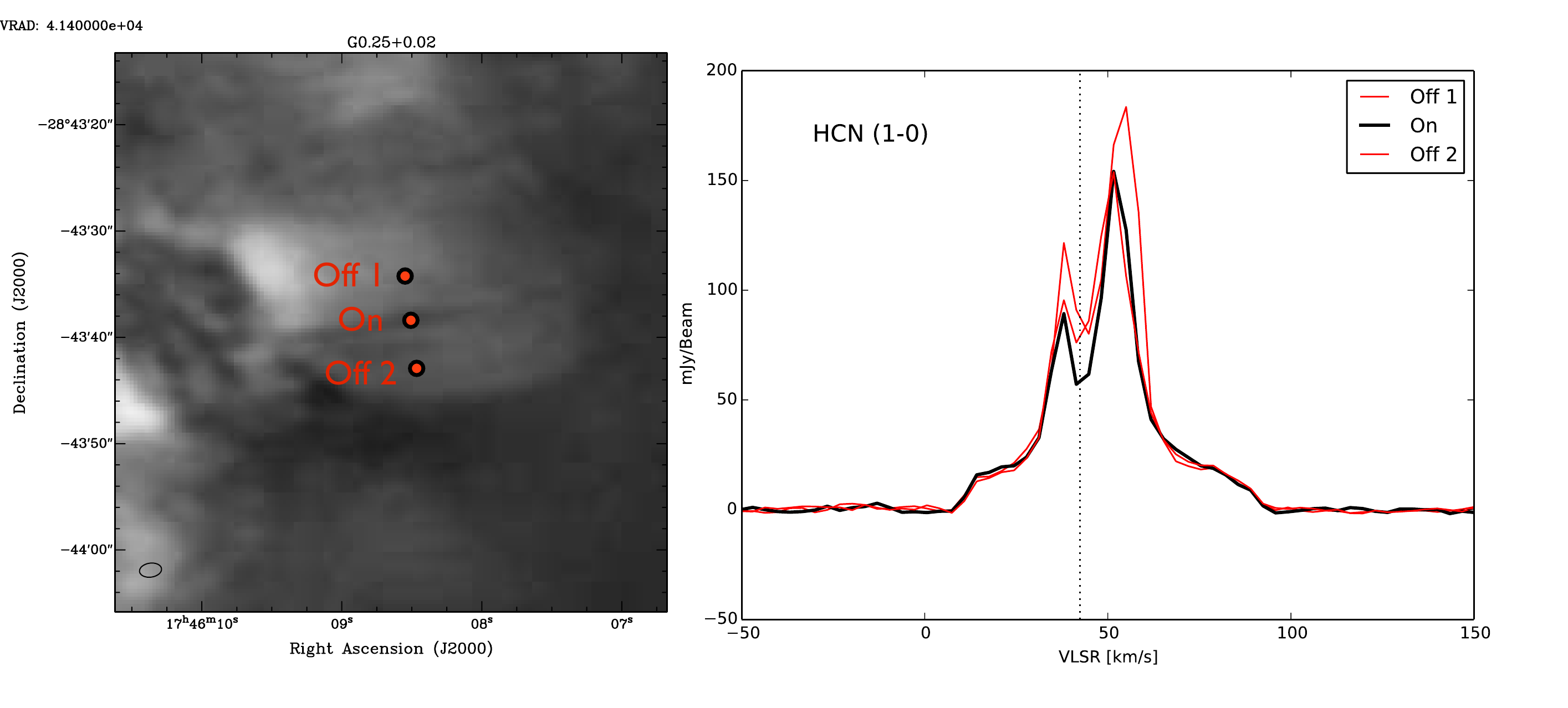}
         }
\caption{ {\it(Left:)} The location of three HCN spectra shown in the Right panel
                 {\it(Right:)}  The HCN spectra shows an example NLA filament  at the location labeled `On'
                  (thick black line) and two off positions (thin red lines). 
                  The dotted vertical line shows the central velocity of the absorption feature.
}
\label{fig11}
\end{figure}

\begin{figure}
\epsscale{1.0}
\center{\includegraphics[width=0.8\textwidth,angle=0]
{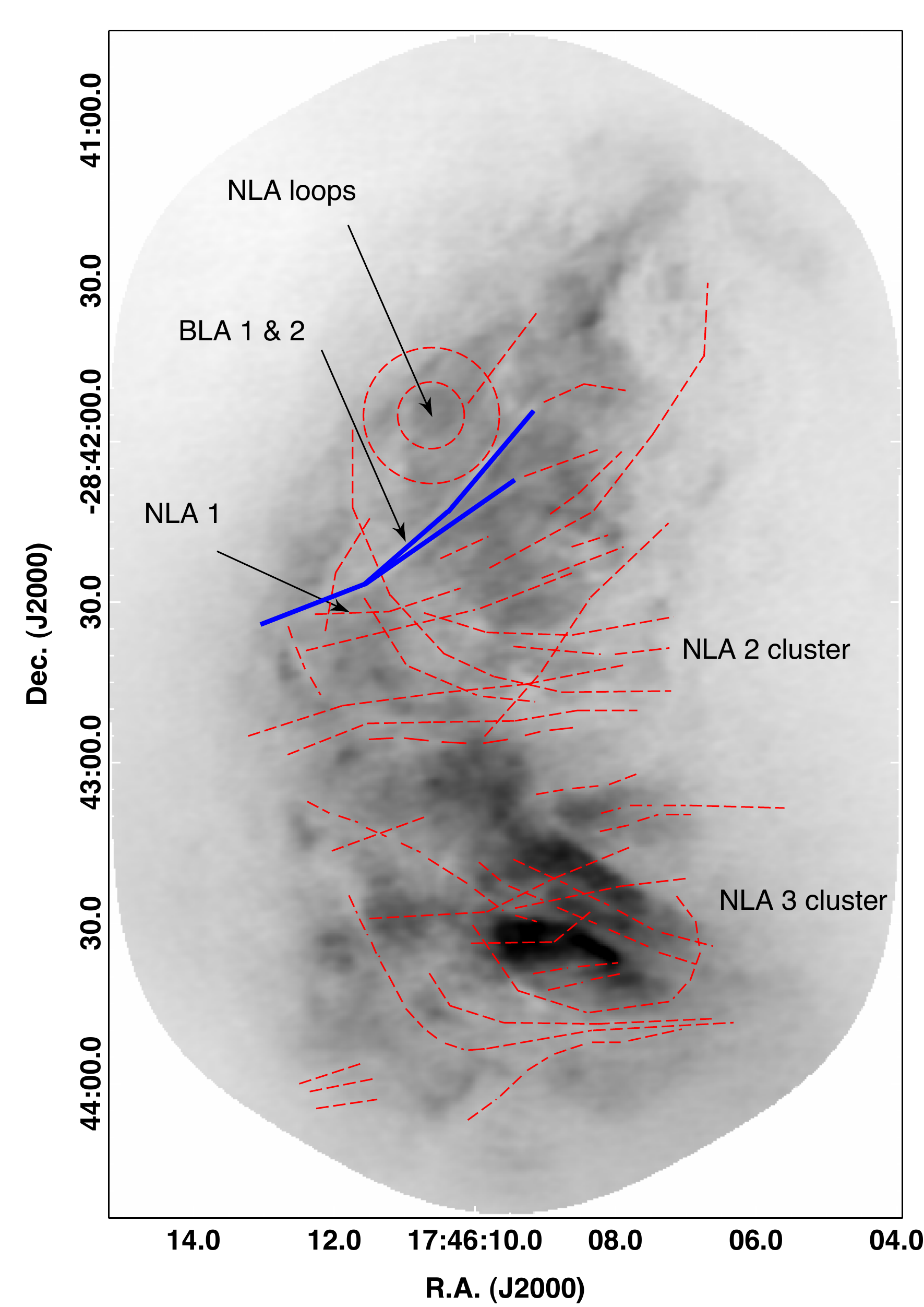}
   }
\caption{
The image shown in Figure 1 with the locations of filaments superimposed.
Thick solid lines show BLA filaments;  thin dashed lines indicate locations of some of
the NLA filaments.  
}
\label{fig12}
\end{figure}

\begin{figure}
\epsscale{1.0}
\center{\includegraphics[width=0.8\textwidth,angle=0]
   {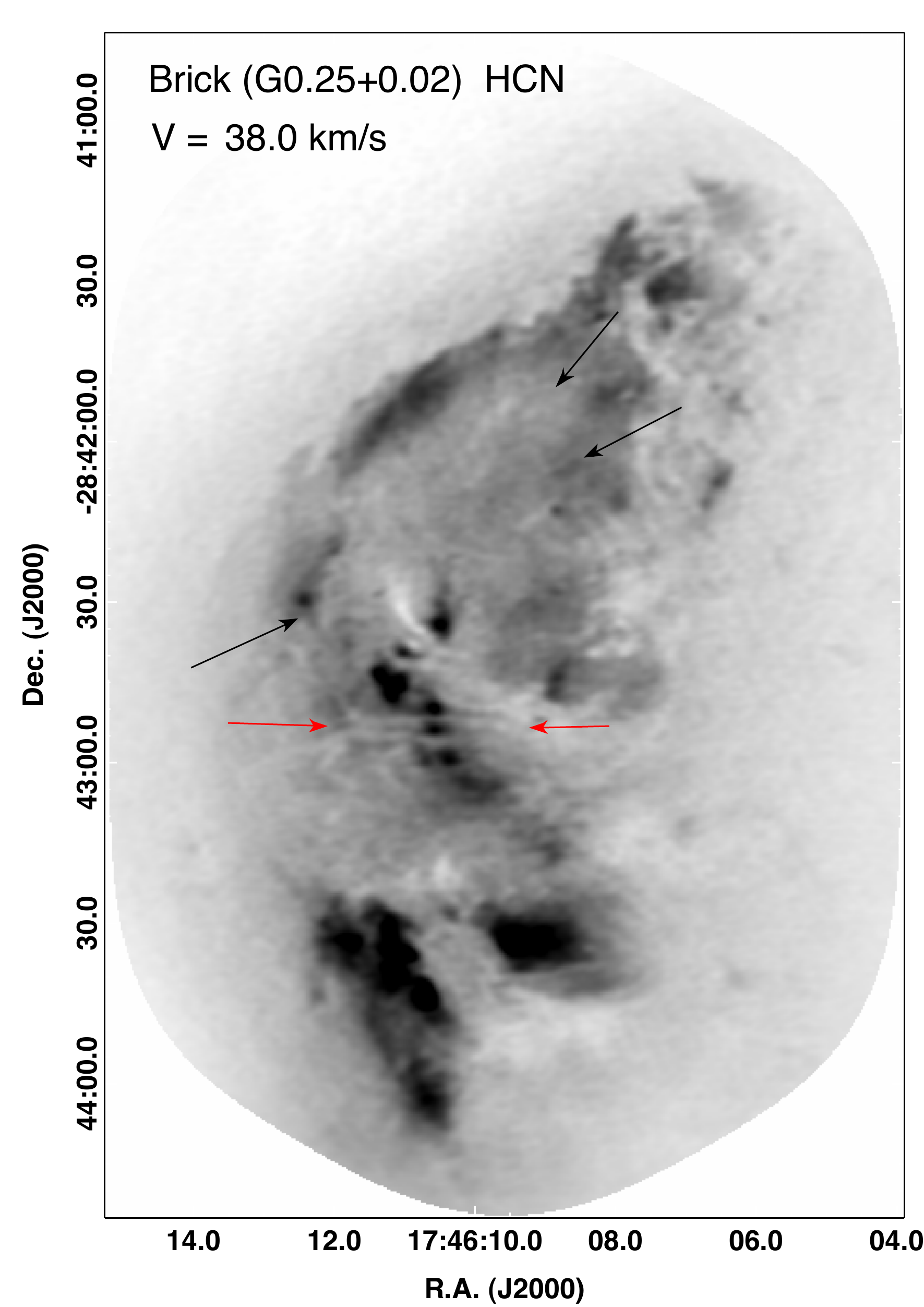}
   }
\caption{
An HCN image at $\rm V_{LSR}$ = 38.0 \kms\ showing narrow-line absorption (NLA)  
filaments.  The prominent broad line absorption (BLA) filaments seen in \HCOP\ are not
seen in this tracer.  The two horizontal arrows mark the locations of several NLA
filaments in the NLA2 cluster which also absorb in HCN.
}
\label{fig13}
\end{figure}

\begin{figure}
\epsscale{1.0}
\center{\includegraphics[width=0.8\textwidth,angle=0]
{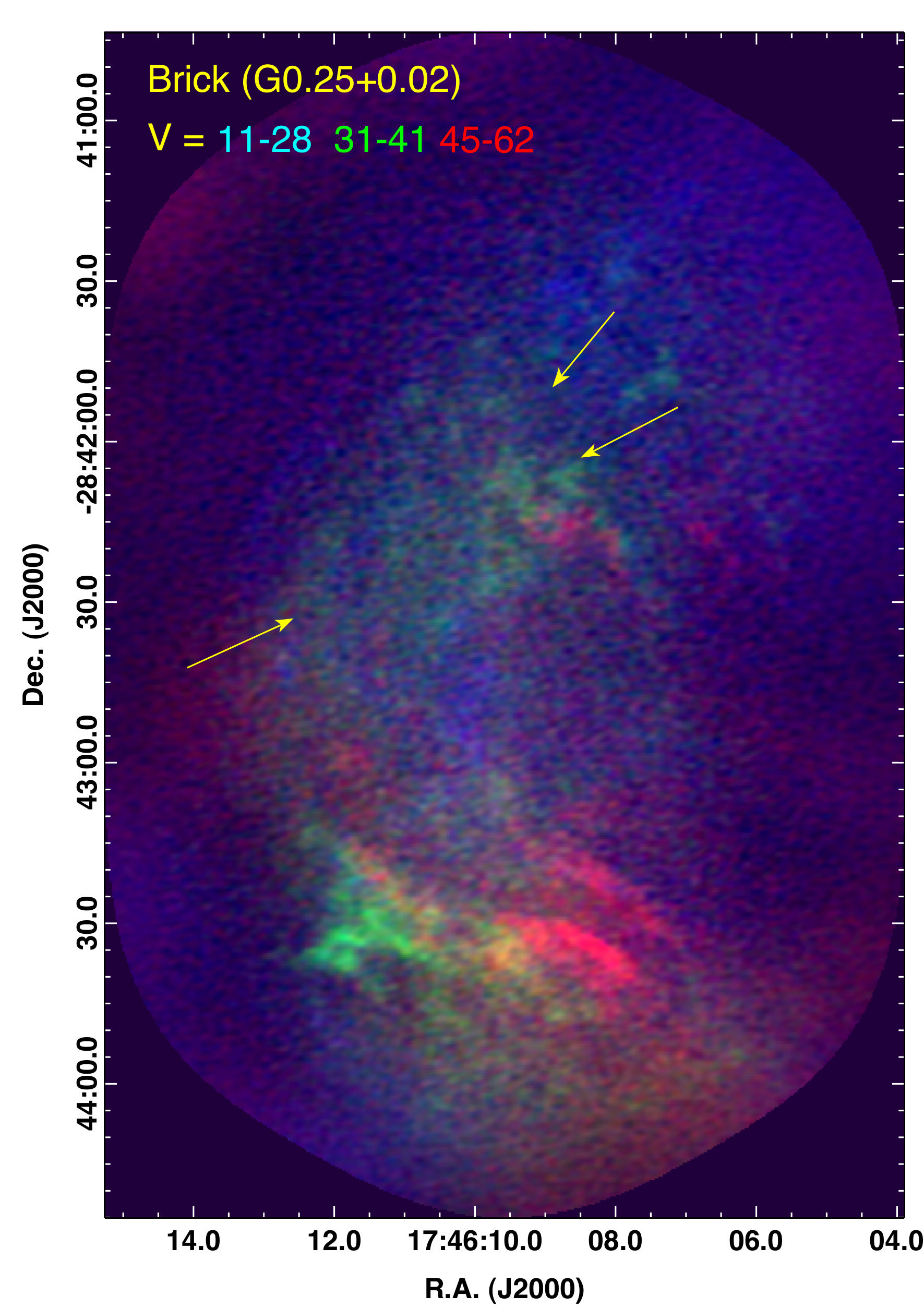}
   }
\caption{
An \hcop\ RGB image showing the bow-shape of the Brick in an optically thin tracer.  
The velocity ranges (in \kms ) for the blue, green, and red are indicated in the figure.
}
\label{fig14}
\end{figure}

\begin{figure}
\epsscale{1.0}
\center{\includegraphics[width=1.1\textwidth,angle=90]
{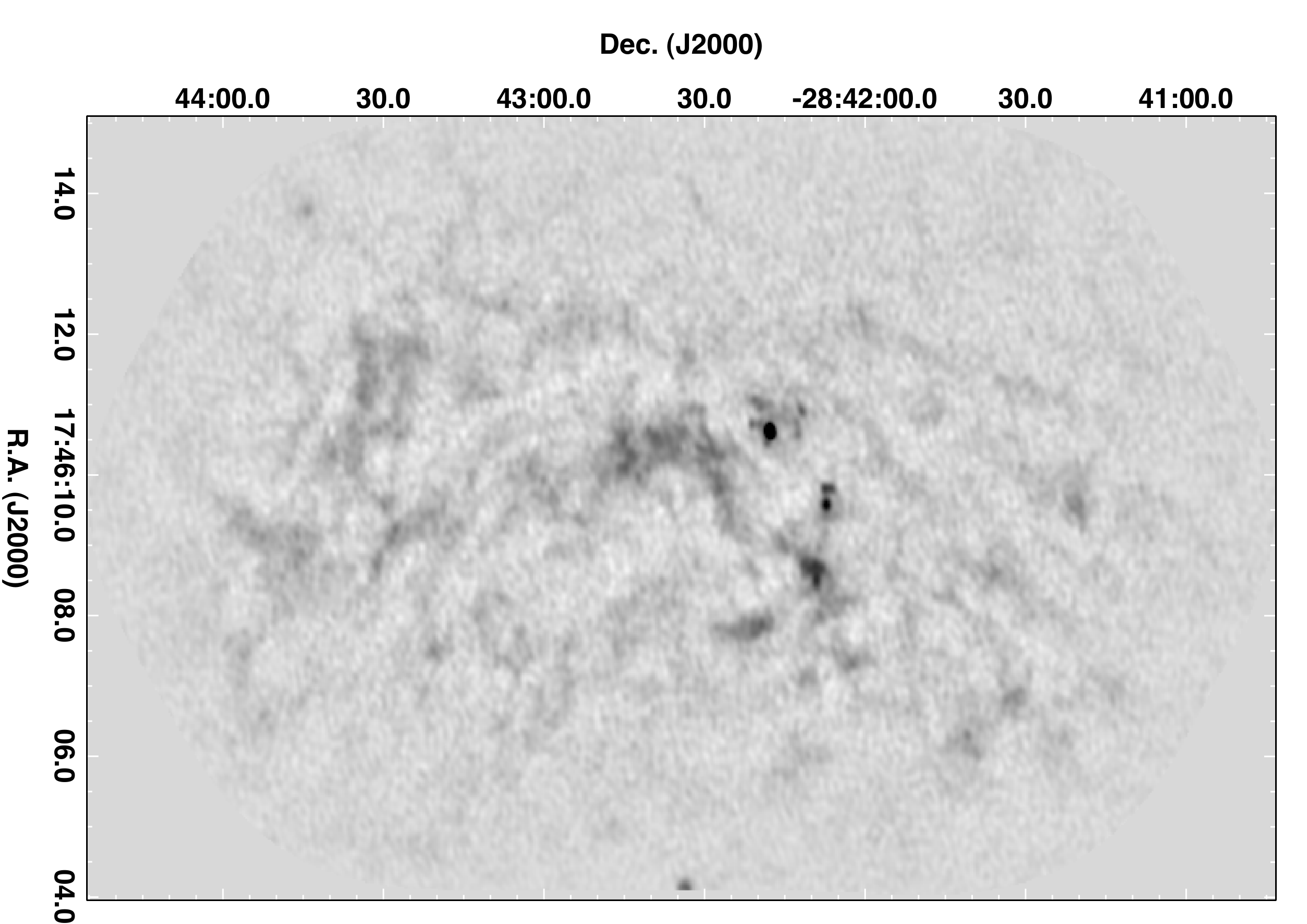}
   }
\caption{
The 3 mm continuum emission from G0.253+0.016.  The bright point
upper left of center is associated with the H$_2$O maser source, located
at J2000 = 17$^h$46$^m$10.62$^s$, $-$28\arcdeg 42\arcmin 17.8\arcsec .  
This image and its interpretation is discussed in detail  in \citet{Rathborne2014_ALMA}.
}
\label{fig15}
\end{figure}

%% file: Table1.tex

\begin{deluxetable}{cccl}
  
\centering

\tablecaption{ 
     \HCOP\ Selected Filaments in Absorption Towards the Brick. 
      \label{table1}}
    
\tablehead{
    \colhead{$\alpha$} & 
    \colhead{$\delta$} & 
    \colhead{V$_{LSR}$} &  
     \colhead{Comments}
       \\

    \colhead{J2000} & 
    \colhead{J2000} & 
     \colhead{km~s$^{-1}$} & 
    \colhead{} 
   
    }
      
\startdata

17:46:09.2	&  -28:41:54	&   21 to 45	& BLA 1  Northwest end \\
17:46:10.6 	&  -28:42:14  	&   21 to 45	&  BLA1  core   \\
17:46:11.8   	&  -28:42:25  	&   21 to 45	&  BLA 1 Southeast end \\
  
17:46:08.9	&  -28:42:05	&   21 to 45 	& BLA2 West end 	\\
17:46:09.9	&  -28:42:12	&   34 to 45 	& BLA2  core 1	\\
17:46:11.1	&  -28:42:22	&   21 to 45	& BLA2  core 2 	\\
17:46:12.9	&  -28:42:34	&  18 to 45	& BLA2 East end 	\\

17:46:10.6	&  -28:41:55	&  11 to 21        & 6\arcsec\ radius loop \\
17:46:10.6	&  -28:41:55	&  11 to 21        & 13\arcsec\ radius loop \\

17:46:10.3	&  -28:42:28	&   31  to 38	& NLA1 West end   	\\				
17:46:11.9	&  -28:42:33	&   31  to 38	& NLA1 East end   \\  

17:46:07.5	&  -28:42:50	&   21 to 45 & NLA2 cluster;  West end   \\
17:46:11.7	&  -28:42:50	&   21 to 45 & NLA2 cluster;  East end   \\

17:46:06.6	&  -28:43:46	&  31 to 48 & NLA3 cluster; West end  \\
17:46:10.6	&  -28:43:30 	&  31 to 48 & NLA3 cluster; East end  \\

\enddata


 \end{deluxetable}

\clearpage

%% file: Table2.tex

\begin{deluxetable}{cccccccccc}
  
\centering

\tablecaption{ 
     RADEX$^1$  calculations for the J=1-0 \HCOP\  transition. 
      \label{table1}}
    
\tablehead{
    \colhead{$log~ n(H_2)$} & 
    \colhead{} & 
    \colhead{T$_{ex}$} & 
    \colhead{} & 
    \colhead{} &  
    \colhead{$\tau$} &
    \colhead{} &
    \colhead{} &
    \colhead{$T_R$} 
     
    }
      
\startdata
$log ~N(HCO^+)$ & 13 & 14 & 15 & 13 & 14 & 15 & 13 & 14 & 15 \\
\hline
2.000	&	2.74	&  2.78 & 	 3.15	&  	0.85	& 8.41 & 72.9	& 0.005 & 0.004 & 0.355 \\
2.477	&	2.76	&  2.87 &	3.78	&	0.84	& 8.10 & 58.1	& 0.016 & 0.118 & 0.909 \\
3.000	&	2.84	&  3.17 & 	5.23 &	0.82	& 7.24 & 37.9	& 0.053 & 0.366 & 2.254 \\
3.477	&	3.07	&  3.83 & 	7.69	&	0.75	& 5.74 & 21.5	& 0.149 & 	0.950 & 4.624 \\
4.000	&	3.84	&  5.49 & 	12.4	&	0.57	& 3.55 & 10.1	& 0.415 & 2.426 & 9.275 \\
4.477	&	6.30	&  9.33 & 	20.1	&	0.30	& 1.69 & 4.60	& 0.849 & 5.078 & 16.77 \\
5.000	&	38.7	&  37.4 & 	41.4	&	0.03	& 0.29 & 1.57	& 1.210 & 8.670 & 30.16 \\
5.477	&	-19	& -34    & 	152	&	-0.04 & -0.2 & 0.31	& 1.002 & 8.681 & 39.68

\enddata

 \tablecomments{
 {\bf[1]}:     The tabulated values assume a  
  background radiation temperature of 2.73 K, 
  a gas  kinetic temperature of 60 K, 
  and a line-width of 10 \kms .  Based on the on-line RADEX tool  available at
  http://www.sron.rug.nl/$\sim$vdtak/radex/radex.php.
       }

 \end{deluxetable}

\clearpage